\documentclass[twocolumn,letterpaper]{aa}
\usepackage{txfonts}
\usepackage{graphicx}
\usepackage{amssymb}
\begin{document}
%
   \title{The WENSS \& Dwingeloo surveys and the Galactic magnetic field}
   \authorrunning{Schnitzeler et al.}

   \author{D.H.F.M. Schnitzeler
          \inst{1}
          \and
          P.Katgert\inst{1}
          \and
	  M.Haverkorn\inst{2,3}
	  \and
	  A.G. de Bruyn\inst{4,5}
          }

   \offprints{D. Schnitzeler, \newline e-mail: schnitze@strw.leidenuniv.nl}

   \institute{Leiden Observatory, Leiden University, P.O. Box 9513, 2300 RA Leiden, The Netherlands\\
              \email{schnitze@strw.leidenuniv.nl}
         \and
              Jansky Fellow, National Radio Astronomy Observatory
         \and 
              Astronomy Department UC-Berkeley, 601 Campbell Hall,
              Berkeley CA 94720, USA
         \and
             ASTRON, P.O. Box 2, 7990 AA Dwingeloo, The Netherlands
         \and 
             Kapteyn Institute, P.O. Box 800, 9700 AV Groningen, The Netherlands
             }

   \date{Received Day Month Year; accepted Day Month Year}

   \abstract{}{}{}{}{} 
   \abstract
    {}
    {We investigate the structure of the Galactic magnetic field in the 2nd Galactic quadrant using radio continuum polarization data from the 325 MHz WENSS (\emph{WE}sterbork \emph{N}orthern \emph{S}ky \emph{S}urvey) survey in combination with earlier single-dish observations.}
    {We study gradients in polarization angle along Galactic longitude and latitude in the region 130\degr\hspace{0.2mm} $\lesssim$ $l$ $\lesssim$ 173\degr\hspace{0.2mm} and $-$5\degr\hspace{0.2mm} $\lesssim$ $b$ $\lesssim$ 31\degr. These gradients were determined with a new method that we developed to efficiently and reliably fit linear gradients to periodic data like polarization angles. Since the WENSS data were obtained with a synthesis array they suffer from the 'missing short spacing' problem. We have tried to repair this by adding an estimate of the large-scale emission based on the single-dish (Dwingeloo) data obtained by Brouw and Spoelstra. Combining all available data we derive a rotation measure map of the area, from which we estimate all 3 components of the magnetic field vector.}
    {In the part of WENSS where large-scale structure in polarized intensity is relatively unimportant, we find that the magnetic field is predominantly perpendicular to the line-of-sight, and parallel to the Galactic plane. The magnetic field components along the line-of-sight and along Galactic latitude have comparable values, and the strength of these components is much smaller than the strength of the total magnetic field. Our observations also cover part of the so-called `fan' region, an area of strong polarized intensity, where large-scale structure is missing from our WENSS data. We tentatively show that Faraday rotation occurring in front of the Perseus arm is causing both the WENSS $RM$ and the $RM$ towards the fan region observed in previous single-dish surveys, and we suggest that the fan is formed by local emission that originates in front of the emission we see in WENSS.}
    {}

   \keywords{Magnetic fields -- Radio continuum: ISM -- Techniques: polarimetric -- ISM: magnetic fields -- ISM: structure -- Techniques: miscellaneous
            }

   \maketitle
%

\section{Introduction}
The diffuse radio emission from our Galaxy provides a powerful tool for studying the various components of the interstellar medium. For example, the well-known all-sky survey by Haslam et al. (\cite{haslam81}) at 408 MHz was used by Beuermann et al. (\cite{beuermann85}) to construct a two-component model of the synchrotron emissivity of our Galaxy. Large parts of the Milky Way have been studied also at other frequencies both in total intensity and in polarized intensity, e.g. Berkhuijsen (\cite{berkhuijsen71}), Reich et al. (\cite{reich97}), Uyaniker et al. (\cite{uyaniker03}), Bingham \& Shakeshaft (\cite{binghamshakeshaft}), Brouw \& Spoelstra (\cite{brouwspoelstra76}).

High-resolution polarization studies came about with the work by Junkes et al. (\cite{junkes87}), and an aperture synthesis array was first used by Wieringa et al. (\cite{wieringa93}). Haverkorn et al. (\cite{haverkorn00}) were to the authors' knowledge the first to use an interferometer to derive information on the magnetic field structure in detail using the diffuse emission of our Galaxy. Presently, the International Galactic Plane Survey (IGPS) project is studying a major part of the Galactic plane up to a couple of degrees on either side and at several frequencies (the polarization data from the IGPS are discussed in Taylor et al. \cite{taylor03} for the Canadian Galactic Plane Survey and by Haverkorn et al. \cite{haverkorn06igps} for the Southern Galactic Plane Survey). 

Equipartition arguments have been used to estimate the strength of the magnetic field (Beck et al. \cite{beck03}). To study the topology of the field various groups are using different techniques. In particular pulsars and extragalactic sources have provided a large amount of information on the large-scale field and on possible reversals in the direction of the field (see e.g. Han et al. \cite{han04}, \cite{han06}, and Brown et al. \cite{brown03}). 
Extragalactic sources sample the entire line-of-sight through the Galaxy, whereas pulsars can tell us about variations along the line-of-sight if we know their position in the Galaxy. Most pulsars are however confined to the Galactic plane, a drawback extragalactic sources do not have.

One major problem with both extragalactic sources and pulsars is that except in parts of the Galactic plane  (Brown et al. \cite{brown03}, Haverkorn et al. \cite{haverkorn06}) the surface density of extragalactic sources and pulsars is low, leaving large gaps between the sampled lines-of-sight. As a result only large-scale structure (LSS) in the magnetic field can be inferred. Extragalactic sources can also have an intrinsic rotation measure, which makes it more difficult to extract the Galactic contribution from the observed rotation measure.

Pulsars have previously been used to provide information on the small-scale magnetic field (Rand \& Kulkarni \cite{randkulkarni}, Han et al. \cite{han04}). The polarized diffuse Galactic radio background can also be used for this purpose, as shown by Haverkorn et al. (\cite{haverkorn04}), who derived the strength of the magnetic field components along the line-of-sight and perpendicular to it, making it possible to study both the small-scale and the large-scale magnetic field. One problem with this method is that the diffuse emission is not very strong, and it can be severely depolarized along the line-of-sight. Correcting for this is complicated by the fact that the amount of emission and depolarization can vary along the line-of-sight. Furthermore the diffuse emission fills the entire telescope beam, which makes modelling more involved than for pulsars and extragalactic sources which are bright point sources.

In this paper we study the large-scale properties of the polarized diffuse Galactic background at 325 MHz using the WENSS survey (Rengelink et al. \cite{rengelink97}). We selected a region spanning about 1000 square degrees between 130\degr\hspace{0.2mm} $\lesssim$ $l$ $\lesssim$ 173\degr\hspace{0.2mm} and $-$5\degr\hspace{0.2mm} $\lesssim$ $b$ $\lesssim$ 31\degr. This area encompasses the Auriga and Horologium regions studied by Haverkorn et al. (\cite{haverkorn03a}, \cite{haverkorn03b}) using multi-frequency data, as well as a large part of the so-called `fan' region, an area bright in polarized intensity, where the magnetic field component in the plane of the sky is very regularly distributed (Brouw \& Spoelstra \cite{brouwspoelstra76} and Spoelstra \cite{spoelstra84}). 

In Sect. \ref{intro data} we describe the WENSS dataset that we use and the reduction of this dataset. Since this dataset was obtained with an interferometer, LSS will be missing from our observations. In Sect. \ref{missing lss} we estimate the contribution of LSS, and we add this estimate to WENSS. We present the results from the original WENSS dataset and that including the LSS estimate in Sect. \ref{secresults}. In Sect. \ref{secgradients} we describe how we analyse the polarization-angle data. We also discuss the robustness of our LSS reconstruction. In Sect. \ref{grad results} we present the results of our analysis in terms of large-scale gradients along Galactic longitude and latitude. In Sect. \ref{comparison to previous results} we compare the gradients in polarization angle that we obtain in areas where no LSS is missing from the interferometer observations to gradients in the rotation measures derived by Spoelstra (\cite{spoelstra84}). This enables us to estimate how reliably gradients in polarization angle can be translated into gradients in $RM$. 
In Sect. \ref{secdiscussion} we discuss the physical picture we can draw from our observations; in particular we attempt to reconstruct all components of the 3D magnetic field vector. In this section we also consider some of the implications of our observations for the nature of the the fan region.







\section{Description of the data}\label{intro data}
\subsection{WENSS}
WENSS, the Westerbork Northern Sky Survey, is a low-frequency radio
survey of extragalactic sources at 325 MHz 
above $\delta=30\degr$\hspace{0.2mm} that also contains a wealth of
diffuse polarization data (Rengelink et al. \cite{rengelink97}). WENSS
has a 5 MHz bandwidth divided over 7 channels, and its sensitivity is 18
mJy (5$\sigma$) at 325 MHz. The survey was carried out using the
Westerbork Synthesis Radio Telescope (WSRT). The WSRT consists of 14
25m dishes, 10 of these are fixed and 4 can be moved along a track,
which improves (u,v) coverage.

\begin{table*}[t]
\centering
\caption[]{Summary of the observations. For each mosaic the dates (yy/mm/dd) and starting times (UT) are shown of the individual 12 hour observing runs that make up the mosaic. Each run is indicated by the length of its shortest baseline. For each mosaic we also show the central coordinates (both equatorial and Galactic), as well as the resolution (FWHM) after applying a Gaussian taper. We used this resolution in further analysis of the data.
}
\begin{tabular}{lcccccc} \hline\hline
\noalign{\smallskip}
Mosaic & WN50\_074 & WN50\_090 & WN66\_045 & WN66\_064 & WN66\_083 & WN66\_102 \\ 
 \noalign{\smallskip}
\hline
 \noalign{\smallskip}
$(\alpha,\delta)_{1950.0}$ &  ($74\degr,50\degr$)  & ($90\degr,50\degr$) & ($45\degr,66\degr$) & ($64\degr,66\degr$) & ($83\degr,66\degr$) & ($102\degr,66\degr$) \\
$(l,b)$ & (157\degr,6\degr) & (163\degr,14\degr) & (136\degr,8\degr) & (142\degr,12\degr) & (146\degr,19\degr) & (149\degr,26\degr) \\
Resolution       & $6.9\arcmin\times9.0\arcmin$  & $6.9\arcmin\times9.0\arcmin$ & $6.9\arcmin\times7.6\arcmin$ & $6.9\arcmin\times7.6\arcmin$ & $6.9\arcmin\times7.6\arcmin$ & $6.9\arcmin\times7.6\arcmin$  \\ 
\noalign{\smallskip}
\hline
\noalign{\smallskip}
 36 m  & 92/01/07    & 92/01/13    & 93/10/18    & 92/01/12    & 92/01/09    & 92/01/08    \\ 
       & 15:24       & 16:06       & 18:45       & 14:26       & 17:36       & 17:14       \\
 48 m  & 91/12/31    & 92/01/06    & 93/10/24    & 92/01/05    & 92/01/02    & 92/01/01    \\ 
       & 15:52       & 16:33       & 18:22       & 14:53       & 16:21       & 17:41       \\
 60 m  & 91/12/24    & 91/12/30    & 93/10/28    & 93/01/10    & 91/12/26    & 91/12/25    \\ 
       & 16:20       & 17:01       & 18:06       & 14:31       & 16:49       & 18:09       \\
 72 m  & 91/12/06    & 91/12/02    & 93/11/01    & 91/12/01    & 91/12/05    & 91/12/04    \\ 
       & 17:41       & 18:51       & 17:50       & 17:11       & 18:12       & 19:32       \\
 84 m  & 91/12/10    & 91/12/16    & 93/11/07    & 91/12/15    & 91/12/21    & 91/12/11    \\ 
       & 17:48       & 17:56       & 17:27       & 16:16       & 17:44       & 19:04       \\
 96 m  & 91/12/17    & 91/12/23    & 94/10/09    & 91/12/22    & 91/12/19    & 91/12/18    \\ 
       & 16:47       & 17:28       & 19:22       & 15:48       & 17:16       & 18:37       \\ 
\noalign{\smallskip}
\hline\hline
\noalign{\smallskip}
\end{tabular}
\label{mosaicinfo}
\end{table*}

To map large parts of the sky in a reasonable amount of time WENSS
uses the mosaicking technique in which the telescope cycles through 78
or 80 pointings (depending of the declination of the mosaic),
integrating each pointing for 20 seconds, and then moving on to the
next. If the pointing centers are on a square grid with a separation
of 1.5\degr\hspace{0.2mm} the off-axis instrumental polarization is
very effectively suppressed (down to the 1\% level - Wieringa et
al. \cite{wieringa93}). In this way each mosaic covers roughly 200
square degrees, and in every night each pointing can be observed about
12 times, yielding visibility data along 12 `spokes' in the (u,v)
plane.

Each WENSS mosaic is formed by combining six 12hr observing runs.  In
each run the telescopes sample the (u,v) plane with a 72m baseline
increment, and different runs have a different shortest baseline
ranging from 36m to 96m in 12m intervals. This results in baselines
from 36m to 2760m. At 325 MHz this gives a maximum resolution of
$54\arcsec\times54\arcsec\csc\delta$ (FWHM). In the analysis of the data we
used a Gaussian taper with a value of 0.25 at a baseline length of
250m to increase the signal-to-noise ratio for more extended
structures. This degrades the resolution to
$6.7\arcmin\times6.7\arcmin\csc\delta$. By adding six 12hr
observing runs the first grating ring is at 4.4\degr\hspace{0.2mm},
i.e. outside the 3\degr$\times$3\degr\hspace{0.2mm} area that we
mapped in the individual pointings from which we construct the
mosaics.
 
The 6 mosaics we selected were mostly observed at night. Observing at
night limits solar interference, and strong ionospheric $RM$ variations
are considerably smaller when the observations are not taking place
during twilight. Information on the mosaics can be found in Table
\ref{mosaicinfo}. For the observations that started in the afternoon
ionospheric effects are probably small since polarized point sources
are still point-like in the data.  After the polarization
calibration described in the next section we regridded the Stokes Q
and U maps from the equatorial to the Galactic coordinate system using
the AIPS task REGRD, after which we calculated the polarized intensity
and polarization angle. Polarization angles were corrected for the
(local) parallactic angle between the equatorial and Galactic north
poles. 

\subsection{Polarization calibration}
Rengelink et al. (\cite{rengelink97}) carried out the total intensity
calibration as part of the WENSS survey. For details about this part
of the data reduction we refer the reader to their paper. Polarization
calibration requires additional reduction steps that we carried out
using the NEWSTAR data reduction package, and we will briefly describe
these here.  For a proper (mathematical) treatment of the different
steps involved in the polarization calibration we refer the reader to
the articles by Hamaker et al. (\cite{ham96}) and Sault et
al. (\cite{sault96}).

Corrections for deviations in the alignment and the ellipticity of the
antenna dipoles, which cause total intensity I to leak into Stokes U
and V, were found using the unpolarized calibrator sources 3C48 and
3C147. The flux scales of these sources is set by the calibrated flux
of 3C286 (26.93 Jy at 325 MHz), determined by Baars et
al. (\cite{baars77}).

Due to an a-priori unknown phase offset between the responses of
the X and Y dipoles there is cross-talk between Stokes U and V. In
principle this phase difference can be calibrated with a polarized
calibrator source, but since such a source was not available for the
WENSS observations we determined the correction by assuming that
Stokes V contains only noise, and minimizing the amount of signal in
V.  For the 5 mosaics observed in January 1992 the correction we found
in this way was $-9\degr$, a demonstration of the stability of the
WSRT on timescales of weeks, whereas for the WN66\_045 mosaic, which
was observed almost 2 years later, it was +26\degr.


Finally we have to correct for the different amounts of ionospheric
Faraday rotation during the different nights in each mosaic and
between mosaics. In each mosaic we found 2-3 bright intrinsically
polarized extragalactic sources, and we used the variations in the
polarization angle of these sources over the different nights to align
the polarization vectors. We did not correct for variations
in ionospheric Faraday rotation within each individual 12hr observing
run; we only took the average polarization angle over each night.  To
correct for polarization-angle offsets between mosaics we compared
polarization angles in the area of overlap between mosaics. The
corrections we found gave a closure error of about 1\degr\hspace{0.2mm} when going
from WN50\_090 to WN50\_074,WN66\_064, WN66\_083 and back to
WN50\_090. The WN66\_045 and WN66\_102 mosaics only have one
neighbouring mosaic, which means that closure errors cannot be
investigated for these mosaics. We assume that the alignment of these
mosaics is of the same quality, even though the width of the
distribution of polarization-angle differences between the WN66\_045
and WN66\_064 mosaics is almost a factor of 2 larger compared to other mosaics.

\section{Estimating large-scale polarization structure missing from WENSS}\label{missing lss}
If an interferometer does not provide visibilities for the central
part in the (u,v) plane LSS in the sky will be
missing from the observations. For the WENSS data structure larger
than about 1.5\degr\hspace{0.2mm} - 2\degr\hspace{0.2mm} is missing. We estimated the
importance of structure on these scales from single-dish data from
Brouw \& Spoelstra (\cite{brouwspoelstra76}, `BS') and Spoelstra
(\cite{spoelstra84}). This (linear) polarization dataset was obtained
with the Dwingeloo 25m telescope at frequencies of 1411, 820, 610, 465 and
408 MHz, beamsizes (FWHM) range from 0.6\degr\hspace{0.2mm} at 1411
MHz to 2.3\degr\hspace{0.2mm} at 408 MHz. These data were however
sampled on an irregular grid, with an average pointing distance of
about 2\degr-3\degr\hspace{0.2mm} below a Galactic latitude of
20\degr, and about 5\degr\hspace{0.2mm} above a latitude of
20\degr. 

Before we can combine the BS and WENSS datasets we have to resample
the BS data onto the WENSS grid, and we have to extrapolate the BS
data from 408 MHz to the WENSS observing frequency of 325 MHz.
These issues will be addressed in the next two subsections. In the
third subsection we discuss how we added the two datasets. In Sect.
\ref{robustness_reconstruction} we will discuss the robustness
of the maps we made of the WENSS data including our estimates of the
missing LSS.

\subsection{Interpolating the Brouw \& Spoelstra measurements}

We addressed the problem of the irregular BS sampling by convolving
the BS datapoints with a Gaussian kernel. The severe undersampling of the BS data sets a lower limit to the kernel width, and a reasonable upper limit can be found by requiring that the interpolated data should not be too smeared out. 
We used 5 different kernels, with FWHM ranging from
3.1\degr\hspace{0.1mm} to 8.7\degr, to interpolate the BS measurements of Stokes Q and U and polarization
percentages at their 5 observing frequencies. In addition we
interpolated their maps of the intrinsic polarization angle of the
emitted radiation and the $RM$ maps they derived. 


\subsection{Extrapolating from 408 MHz to 325 MHz}\label{extrapolate}
The second problem is how to get from the BS measurements, with a lowest observing frequency of 408 MHz, to the 325 MHz of the WENSS data. 
To predict the polarized intensity at 325 MHz, $P_{\mathrm{pred,325}}$, we start with the observed polarized intensity at 408 MHz, $P_{\mathrm{obs,408}}$, convert it to total intensity and scale it up to 325 MHz using a power-law, and finally convert this back to polarized intensity at 325 MHz:

 \begin{equation}
 P_{\mathrm{pred,325}} = P_{\mathrm{obs,408}}\times(408/325)^{0.7}\times pp_{\mathrm{frac}} 
 \label{p_pred} 
 \end{equation}


\noindent
where $pp_{\mathrm{frac}}$\hspace{1mm} is defined as

 \begin{equation}
  pp_{\mathrm{frac}} \equiv \frac{ \mbox{polarization \% at 325 MHz} }{ \mbox{polarization \% at 408 MHz} } 
 \label{pp_frac}
 \end{equation}

\noindent 
For the polarization angle at 325 MHz, $\phi_{\mathrm{pred,325}}$, we calculate the amount of Faraday rotation from:

 \begin{equation}
  \phi_{\mathrm{pred,325}} = \phi_{0,\mathrm{BS}} + RM_{\mathrm{BS}}\lambda_{(\mathrm{325 MHz})}^2 
 \label{phi_pred}
 \end{equation}

\noindent
where $\lambda_{(\mathrm{325 MHz})}$ is the observing
wavelength corresponding to 325 MHz. The $\phi_{\mbox{\scriptsize
pred}}$ are wrapped back to [$-$90\degr,90\degr]. For these
predictions we use the intrinsic polarization angle of the emitted
radiation $\phi_0$ and $RM$ as derived in Spoelstra
\cite{spoelstra84} (`$\phi_{0,\mathrm{BS}}$' and
`$RM_{\mathrm{BS}}$'), together with the power-law index
they adopt for the brightness temperature, 2.7 (which gives the 0.7 in
Eqn. \ref{p_pred} since the 2.7 is used for scaling brightness
temperatures, and we convert total intensities), and the polarization
fractions they derive using this power-law at the different
frequencies.
Eqn. \ref{phi_pred} requires an interpolated map of polarization angles, but 
interpolation cannot take the periodicity of the data into account. We tried to minimize this effect by interpolating 
$\phi_{0,\mathrm{BS}}$ data, which showed less $\pm$90\degr\hspace{0.2mm} transitions than other 
polarization-angle data.

For the actual prediction of the large-scale Stokes Q and U at 325 MHz
from the 408 MHz observations we must assume a polarization percentage
at 325 MHz. From Fig. 6 in Brouw \& Spoelstra (\cite{brouwspoelstra76})
it is clear that the average polarization percentages decrease
slightly between 610 MHz and 408 MHz for b=0\degr
$\rightarrow$+20\degr. By using the same polarization percentage at
325 MHz and 408 MHz (at each frequency one value for the entire
mosaic) we can be sure that we are not underestimating the predicted
polarized intensity of the BS data at 325 MHz or, equivalently, the
importance of LSS missing from WENSS.

We checked how well this scheme works by using the BS polarization
data and interpolated $pp_{\mathrm{frac}}$ maps at 1411, 820
and 610 MHz to predict the observed polarized intensity at 820, 610
and 408 MHz respectively.
If we do not use a pixel-to-pixel correction for the variation in
polarization percentage but just one correction factor for the entire
mosaic,
we predict on average a polarized intensity that is about the same as the
observed intensity. The spread (standard deviation) in the ratio of
the predicted vs. the observed polarized intensities (about 0.2) is
very similar to that obtained with a pixel-to-pixel correction for
variations in polarization fraction, except at 408 MHz where it is
0.5. The equivalent width at 408 MHz when the pixel-to-pixel
correction is used is still about 0.3. This increased spread can be
explained by varying amounts of Faraday rotation between pixels, which
becomes more important at lower frequencies. 
If part of the emission is actually thermal Bremsstrahlung,
we would be overestimating the amount of emission at lower frequencies
by using a synchrotron power-law with spectral index 2.7. Apparently
this problem is not very important, given the results of our
comparison between the predicted and observed polarized intensities.

From these results we conclude that our recipe to predict polarized intensity
works sufficiently well that we can attempt to predict polarization
data at 325 MHz from the BS 408 MHz data.

\subsection{Adding the LSS estimate to WENSS}\label{addingwensswensslss}
\begin{figure*}[t]
\resizebox{\hsize}{!}{\includegraphics{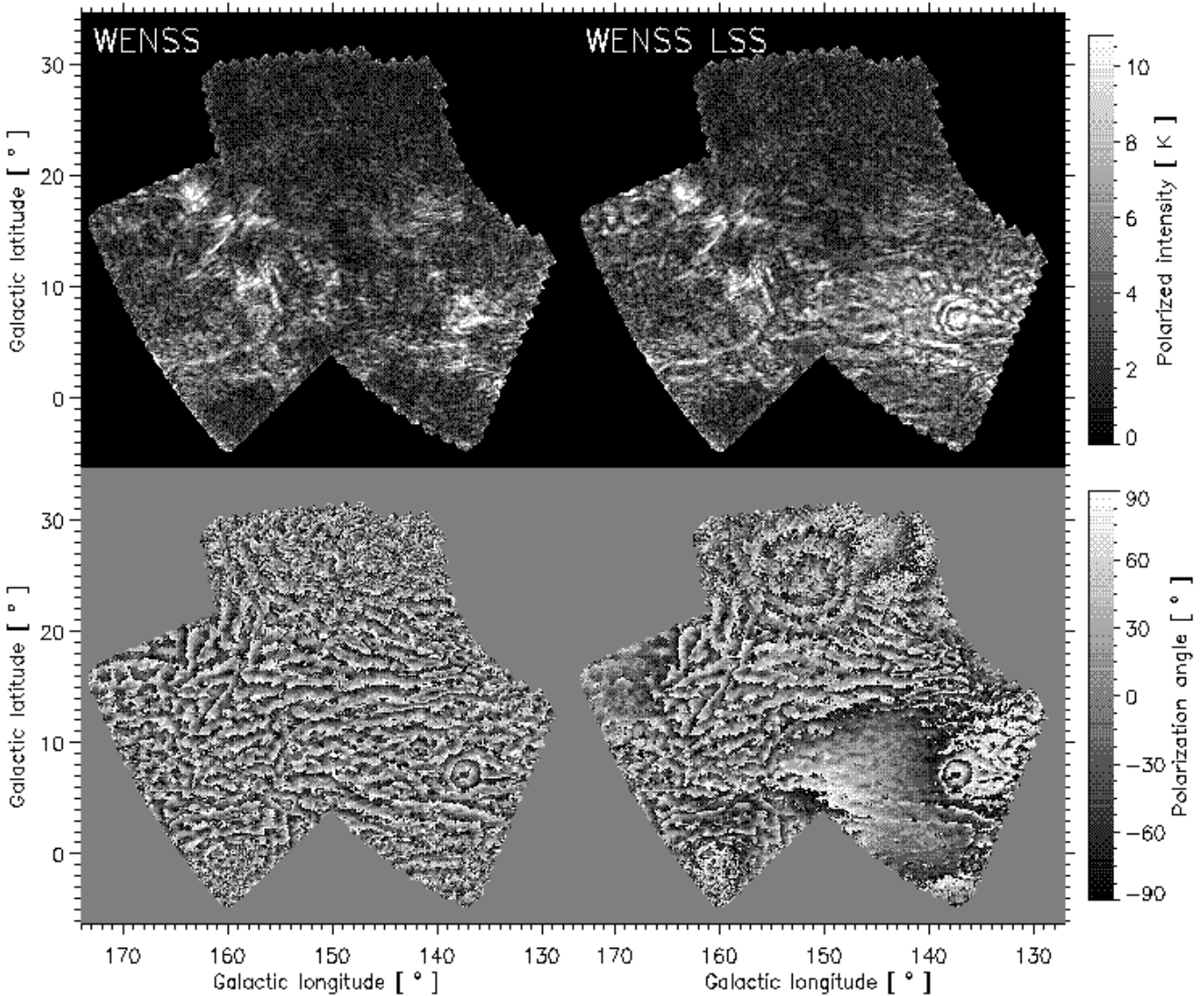}}
\caption{Polarized intensity (top row) and polarization angle (bottom row) mosaics for WENSS (left column) and WENSS including our reconstruction of the missing LSS (`WENSS LSS'; right column). The beamsize of the observations (not indicated) is 6.7'$\times$6.7'$\csc\delta$ (FWHM). Properties of the reconstruction (see the text): width of the convolution kernel is 5.7\degr\hspace{0.2mm} FWHM, $pp_{\mathrm{frac}}$=1, $\Delta\phi$=0\degr. The polarized intensity scale in both mosaics is the same.
The nested rings in polarization angle in WENSS LSS at Galactic latitudes $b>20$\degr\hspace{0.2mm} is an artefact due to a single large $RM$ in the BS data surrounded by much smaller $RM$. 
}
\label{6p_phi_wenss_wensslss}
\end{figure*}

For an excellent review of different methods to add missing
LSS to interferometer observations see Stanimirovic
(\cite{stanimirovic}). We used the `linear combination approach',
where the Stokes Q and U maps from WENSS are added to the
reconstructions of Stokes Q and U from the BS data. However, as we
could not cross-calibrate the WENSS and BS data the intensity scales
are not guaranteed to be identical. A potentially more serious problem
is that the polarization vectors in WENSS and BS may have different
zero-points. In Sect.  \ref{robustness_reconstruction} we will
investigate the effects of such errors.




In the remainder of this paper we will us the term `WENSS LSS' when
we refer to the WENSS dataset complemented by a BS reconstruction of
the missing large-scale Q and U structure at 325 MHz.

\section{Discussion of the maps}\label{secresults}

In Fig. \ref{6p_phi_wenss_wensslss} we show the polarized intensity
(in Kelvin) and polarization angle for WENSS and WENSS LSS
respectively. Two of the brightest features in both WENSS and WENSS
LSS, viz. the V-shaped feature at ($l,b$) = (161\degr, 16\degr) and
the ring-like structure at (137\degr, 7\degr), were analysed in detail
by Haverkorn et al. (\cite{haverkorn03a}, \cite{haverkorn03b}) using
multi-frequency observations.
The most striking feature in Fig. \ref{6p_phi_wenss_wensslss} is the
clear stratification of polarization angles in Galactic latitude. In
polarized intensity there are also linear features aligned with the
Galactic plane, but these are not as clear as the structures in
polarization angle.

After adding the BS data to WENSS the ring-like structure that was
already clearly visible in polarization angle now also becomes more
conspicuously ring-like in polarized intensity.

In WENSS LSS there is an area between 136\degr\hspace{0.2mm} $\lesssim l \lesssim$
152\degr\hspace{0.2mm} and $-$3\degr\hspace{0.2mm} $\lesssim b \lesssim$ 13\degr\hspace{0.2mm} where the polarization angles
apparently show no clear stratification with Galactic latitude. This
region is part of what has previously been referred to by some as the
`fan' region, a large area of high polarized intensity with a strongly
aligned projected magnetic field (see e.g. Figs. 6a to 6e in Brouw \&
Spoelstra \cite{brouwspoelstra76} and Fig. 3a in Spoelstra
\cite{spoelstra84}).
Where the BS signal is strong (in the upper left and lower right of
the maps) there are clear differences between WENSS and WENSS LSS. In
some parts this structure in BS seems to erase the stratification of
polarization angle we noted in the previous paragraph, but in large
parts of WENSS LSS the stratification is still very clear. We will
return to this point in Sect. \ref{secdiscussion}.  The nested imprint
in polarization angle in WENSS LSS above $b=20$\degr\hspace{0.2mm} is
caused by a single line-of-sight in the BS data that has a large $RM$,
surrounded by much smaller $RM$. This creates steep gradients in $RM$
which by Eqn. \ref{rm_grad} are translated into steep polarization
angle gradients in the reconstructed BS data.  There also appears to
be a general change in scale of the polarization features above
$b=20$\degr\hspace{0.2mm} in WENSS. In the WN66\_102 mosaic, which is
the main source of data above $b\gtrsim20$\degr, an unusually large
fraction of the (u,v) datapoints at the shortest baselines had to be
flagged. This could be partly responsible for the observed change in
structure. However, flagging of these data in other mosaics did not
significantly alter the maps.

\section{Quantifying the large-scale stratification of polarization angle: technique}\label{secgradients}
The striking stratification of polarization angle in Fig.
\ref{6p_phi_wenss_wensslss} suggests that there is a clear gradient in
the direction of Galactic latitude, with only a relatively small
variation in Galactic longitude, on which small-scale structure is
superimposed. In order to quantify this large-scale gradient it is
necessary to filter out these small-scale modulations, and also to
correctly handle the 180\degr\hspace{0.2mm} ambiguity of the
polarization angle.


Resolving the 180\degr\hspace{0.2mm} ambiguity by minimizing the
difference in polarization angles of consecutive datapoints is
sensitive to small-scale structure and noise. 
By using longer intervals in longitude or latitude we can fit the
large-scale behaviour in polarization angle without being influenced
too much by small-scale structure, and at the same time resolve the
180\degr\hspace{0.2mm} ambiguity for all fitted datapoints together.

\begin{figure}[t]
\resizebox{\hsize}{!}{\includegraphics{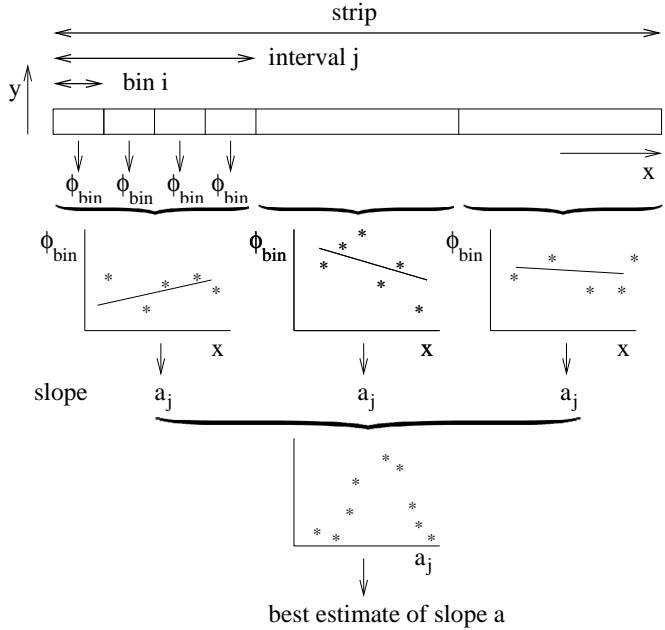}}
\caption{Overview of how we analyse our polarization data. First we derive the average polarization angle in each bin (polarized intensity weighted), $\phi_{\mathrm{bin}}$. Bins are aligned with the `x' and `y' Galactic coordinates. We then fit a gradient to all $\phi_{\mathrm{bin}}$ in an interval, taking into account the periodic nature of the polarization angles, and we repeat this for all intervals in a strip along Galactic longitude or latitude.  The gradients can then be used to investigate the average gradient per strip, as shown in the third row.}
\label{overzicht_gradientfitting}
\end{figure}

\subsection{Determination}\label{secdetermination}
The different steps in our analysis of the polarization-angle data are
illustrated in Fig. \ref{overzicht_gradientfitting}.  To determine the
large-scale gradients we first calculate a weighted average of the
polarization angles over an area of a fixed size (`bin', the average
itself is indicated by `$\phi_{\mbox{\scriptsize bin}}$' in
Fig. \ref{overzicht_gradientfitting}). For strips along Galactic
latitude these bins are 28\arcmin$\times$7\arcmin\hspace{0.2mm} in
size (about 4$\times$1 WENSS beams, taking into account the
orientation of the beam in the Galactic coordinate frame) and for
strips along Galactic longitude
14\arcmin$\times$13\arcmin\hspace{0.2mm} (or about 2$\times$2 WENSS
beams). In this way all bins contain about the same amount of beams,
and the different shapes of the bins reflect the difference in
steepness between gradients along Galactic longitude and latitude. If
the spread of the polarization angles within the bin is larger than
20\degr\hspace{0.2mm} the bin is discarded.  Furthermore if a cut is
placed on e.g. the signal-to-noise ratio of the pixels in a bin, at
least 50\% of the pixels should be usable (not flagged), otherwise the
bin will also be discarded.


We then make a linear fit to all usable $\phi_{\mbox{\scriptsize
bin}}$ in an interval of a specified length, as shown in the second
row of Fig. \ref{overzicht_gradientfitting}. We only made fits to
intervals that contain at least 6 usable bins.

The fits were made using a standard $\chi^2$ minimization.
Instead of determining for every possible combination of gradient and
offset which 180\degr\hspace{0.2mm} `flips' of the datapoints minimize
the $\chi^2$ for that combination it is also possible to determine
which configurations of `flips' are allowed by the data. The best fit to each configuration can then be found using a standard $\chi^2$ minimization, and the configuration with the lowest $\chi^2$ value will give the overall best fit. This method can be shown to be more reliable and
time-efficient than an approach that `probes' which gradient fits the
data best. See appendix A for details.  This technique can also be
extended to situations where the horizontal distance between
datapoints varies, as is the case when polarization-angle observations
at different wavelengths are used to derive $RM$.

For every strip we also fitted gradients that are shifted by half an
interval length to prevent losing information (Nyquist sampling). In
the remainder we include the gradients in these shifted intervals in
our analysis.


\subsection{Robustness of the gradients including the LSS estimate} \label{robustness_reconstruction}
When combining WENSS with our reconstruction of the large-scale
structure there are several free parameters that needed to be chosen:
the FWHM of the convolution kernel for the BS data, the possible
misalignment between the WENSS and BS polarization vectors
($\Delta\phi$) and differences in the amplitude scales of the WENSS and
the BS data (which can be absorbed into differences in
$pp_{\mathrm{frac}}$). We have investigated the effect of
varying these parameters by comparing histograms of polarization-angle
gradients. In the remainder of this section we will consider only
gradients along Galactic latitude and we test for only one interval
length of 1.4\degr. 
As there is much more structure in the
polarization-angle gradients along Galactic latitude as compared to
gradients along Galactic longitude, gradients along Galactic latitude
are more sensitive to changes in the parameters we are investigating. 
An interval length of 1.4\degr\hspace{0.2mm} gives the
largest number of usable gradients in our dataset. Since at least 6
bins per interval must be usable to fit a gradient, the number of
gradients fitted over short interval lengths will be limited.  On the
other hand for an area of a given size the number of intervals that
can be fitted in that area decreases when the interval length goes up.

Note that we only investigate the \emph {statistical} properties of
the polarization angles that we find after adding the large-scale
structure. We do not claim that individual pixels will have the
correct polarized intensity or polarization angle.

Taking $\Delta\phi$=0\degr\hspace{0.2mm} we found no significant
differences in the distributions of the polarization-angle gradients
for all widths of the convolution kernel $\gtrsim$
5\degr\hspace{0.2mm} (FWHM). Smaller widths are simply too narrow to
smoothly interpolate the undersampled BS datapoints, leading to
artifacts.
As there are hardly any differences for the different widths of the
interpolation kernel we will be using the reconstruction made using
the FWHM=5.7\degr\hspace{0.2mm} kernel for further analysis. Since the
WENSS data is missing information on scales $\gtrsim$ 1.5\degr\hspace{0.2mm} -
2\degr\hspace{0.2mm} this means that we do not have information on
scales from about 1.5\degr\hspace{0.2mm} to about 5\degr\hspace{0.2mm}
after adding the BS reconstruction.

Since the WENSS and BS datasets do not overlap in the (u,v) plane it is not
possible to determine if there is a misalignment between the WENSS and
BS polarization vectors. We applied polarization-angle offsets of
30\degr, 60\degr, 90\degr, 120\degr\hspace{0.2mm} and 150\degr\hspace{0.2mm} to the BS reconstruction
with a kernel width of 5.7\degr\hspace{0.2mm} before adding it to WENSS, and found no
significant change in the gradient distribution. We therefore keep $\Delta\phi$
fixed at 0\degr.


Finally we considered the effects of differences in the intensity
scales between WENSS and the reconstructed BS data. These consist of
two contributions, the assumed polarization percentage at 325 MHz,
which is important for $pp_{\mathrm{frac}}$\hspace{1mm}
(cf. Eqn. \ref{p_pred}), and differences in the flux scales
between the WENSS and BS datasets (mentioned in Sect.
\ref{addingwensswensslss}).
         
\begin{figure}[t]
\resizebox{\hsize}{!}{\includegraphics{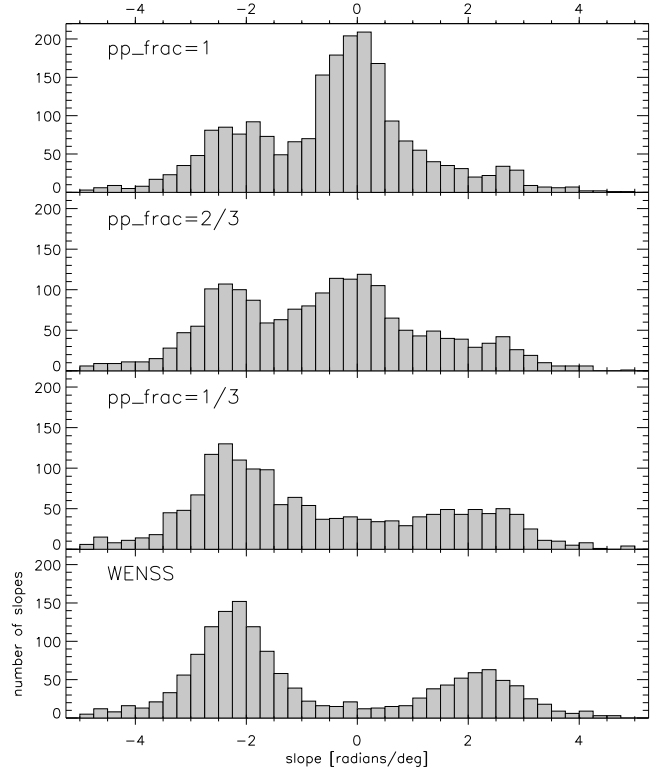}}
\caption{Histograms of gradients fitted to the WENSS LSS data and to
WENSS. The top panels show the result for a reconstruction of WENSS
LSS using a $pp_\mathrm{frac}$ (defined in Eqn. \ref{pp_frac}) of 1,
2/3 and 1/3 respectively. In the bottom panel we plot the distribution
of the gradients fitted to the WENSS data without the LSS
reconstruction. In all cases the width of the convolution kernel used
to reconstruct the LSS is 5.7\degr\hspace{0.2mm}
(FWHM), $\Delta\phi$=0\degr, and the interval length over which the
gradients were fitted is 1.4\degr.
}
\label{hist_slopes_run44_run45_run92}
\end{figure}

In our reconstruction of the LSS missing from WENSS
we assumed that the ratio of the polarization fractions at 325 MHz and
408 MHz (Eqn. \ref{pp_frac}) was 1. This ratio is more likely less
than 1, maybe as low as 2/3, which would imply that there is less
LSS missing from WENSS than we assumed. In the top
two panels of Fig. \ref{hist_slopes_run44_run45_run92} we show the
gradient distributions for these two values of $pp_\mathrm{frac}$. The only major difference between the histograms is the
fraction of gradients that are in the peak around 0 radians/deg. The
second peak around $-$2.3 radians/deg hardly changes.  In Sect.
\ref{comparing WENSS LSS to WENSS} we identify these peaks and explain
the difference for the two values of $pp_\mathrm{frac}$. For now it suffices to say that this difference does not
influence our results.

Since the influence of calibration differences between the WENSS and
BS flux scales is probably smaller than the range of the two values of
$pp_{\mathrm{frac}}$ we probed, we did not consider this
difference in flux scale as a separate issue.


 
\subsection{Dependence on the value of $pp_\mathrm{frac}$}\label{comparing WENSS LSS to WENSS}

In Fig. \ref{hist_slopes_run44_run45_run92} we plotted the distributions of the gradients fitted to the WENSS LSS dataset for $pp_{\mathrm{frac}}$=1, 2/3, 1/3 and for the original WENSS distribution ($pp_{\mathrm{frac}}$=0). In this way the gradual change in the distribution between the different reconstructions becomes clear.


Gradients in WENSS are found in two peaks at +2.3 radians/deg and
$-$2.3 radians/deg. Since WENSS LSS is constructed as the vector sum
of the WENSS and BS datasets, the relative strength of the BS and
WENSS polarization vectors determines if the gradients in WENSS LSS
show the clear bimodal behaviour of the original WENSS dataset. In
this way the negative mode and tentative positive mode in the
distribution of gradients in WENSS LSS in
Fig. \ref{hist_slopes_run44_run45_run92} can be identified as the two
modes of the bimodal WENSS distribution shown in the bottom panel of
Fig. \ref{hist_slopes_run44_run45_run92}. 

When $pp_\mathrm{frac}$ increases, the BS signal becomes stronger
compared to the WENSS signal, and the area where the BS signal is
significantly stronger than the WENSS signal also increases. Both
factors will `convert' gradients from the steep gradients in the peaks
in the bimodal WENSS distribution to flatter BS gradients that are
found in the peak around 0 radians/deg.

\subsection{Dependence on interval length}\label{vertical strips dependence on interval length}
In Sect. \ref{robustness_reconstruction} we discussed the distribution
of the fitted gradients along Galactic latitude
$\partial\phi_{\mathrm{pol}}/\partial b$ for an interval length of
1.4\degr. The distributions in the top panels of Fig.
\ref{hist_slopes_run44_run45_run92} are clearly bimodal, with a
central dominant peak around 0 radians/deg and a second clear
peak at about $-$\nolinebreak2.3 radians/deg. Furthermore there is
tentative evidence for a plateau or small peak at around +2.5
radians/deg. For the shortest 2 interval lengths we probed
(0.7\degr\hspace{0.2mm} and 1.1\degr\hspace{0.2mm}) only the central
peak is visible, with a plateau extending to about
$-$2.3\degr/bin. This plateau turns into a clear peak for intervals
longer than 1.4\degr. At longer intervals (2.1\degr, 2.8\degr,
4.2\degr\hspace{0.2mm} and 5.6\degr) the width of the central peak decreases and the
positive peak at +2.5 radians/deg becomes more prominent. For interval
lengths of 4.2\degr\hspace{0.2mm} and 5.6\degr\hspace{0.2mm} the
number of gradients in the central peak decreases rapidly.

\begin{figure}[t]
\resizebox{\hsize}{!}{\includegraphics{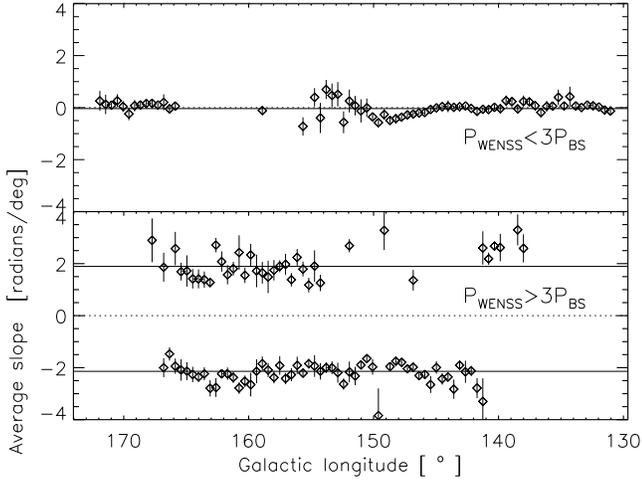}}
\caption{Average gradient per strip along Galactic latitude as a
function of Galactic longitude. In the top panel we plotted gradients
fitted to datapoints where the polarized intensity from WENSS is less
than 3 times the polarized intensity in BS, in the bottom panel we
plotted gradients fitted to the remaining datapoints. In the lower
panel we show the average gradient per strip for the positive and
negative modes separately. We only plotted strips that had at least 4
usable gradients, in the bottom panel each mode has at least 4 usable
gradients. The errorbars are calculated from the spread (standard
deviation) of the gradients in a strip, and are at the 1$\sigma$
level. The weighted average of the datapoints is indicated by a solid
line, and in the bottom panel the weighted average was calculated for
the positive and negative modes separately. The width of the
interpolation kernel was 5.7\degr\hspace{0.2mm} FWHM,
$pp_{\mathrm{frac}}$=1, and the interval length of the
fitted gradients is 1.4\degr.  }
\label{gemhellingverticaal}
\end{figure}

\begin{figure}[t]
\resizebox{\hsize}{!}{\includegraphics{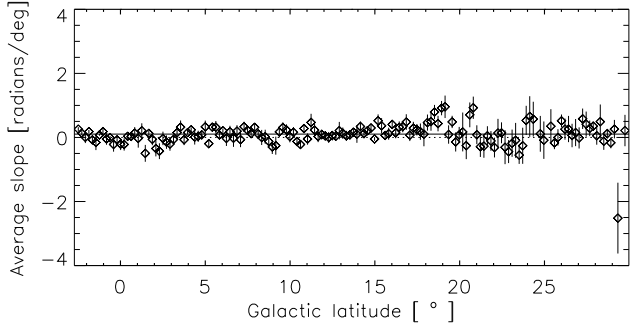}}
\caption{Average gradient per strip along Galactic longitude, as
a function of Galactic latitude, plotted on the same vertical scale
as in Fig.  \ref{gemhellingverticaal}. We only considered strips that
have a least 4 usable gradients. The errorbars are calculated from the
spread (standard deviation) of the gradients in a strip, and are at
the 1$\sigma$ level.  FWHM of the kernel used in the reconstruction is
5.7\degr, $pp_\mathrm{frac}$=1, interval length over which the
intervals are fitted is 2.8\degr.  }
\label{slopeshor}
\end{figure}

The decreasing width of the peak of gradients around 0 radians/deg
when going to longer intervals can be explained by the fact that
longer intervals will span positive and negative gradients on shorter
lengthscales, and these `subgradients' will partially cancel each
other. At the same time the bimodal WENSS distribution is found for
all interval lengths in the WENSS-only data. These effects combined
account for the positive WENSS mode becoming more pronounced when
going to longer intervals.



The average $\chi^2_\mathrm{red}$ of the fitted gradients
increases when we use longer intervals, going from $\lesssim$ 4 for
interval lengths of 0.7\degr\hspace{0.2mm} to $>$ 20 for intervals of 5.6\degr. This
is due to structure in polarization angle on scales smaller than the
length of the fitted gradient.

\section{Quantifying the large-scale stratification of polarization angle: results}\label{grad results}
\subsection{Gradients along Galactic latitude}\label{gradlat}\label{avlon}
Fig. \ref{gemhellingverticaal} shows the average polarization-angle gradients along Galactic latitude ${\partial\phi_{\mathrm{pol}}}/{\partial b}$ for each Galactic longitude in our sample, where we separately fitted gradients to datapoints where the ratio of polarized intensities P$_{\mathrm{WENSS}}$:P$_{\mathrm{BS}}$ is $<3$ or $>3$ (top and bottom panels respectively).
This neatly separates the two distribtutions of slopes. In the bottom panel of Fig. \ref{gemhellingverticaal} we furthermore calculate the average gradient per strip separately for the positive and negative modes of gradients. Gradients in the top panel do not show this bimodal distribution. The solid line shows the weighted average over all Galactic longitudes; in the bottom panel the weighted average was calculated for the positive and negative modes separately. In both figures we only considered strips that have at least 4 usable gradients, a minimum number for an average gradient per strip to be statistically reasonable.

From Fig. \ref{gemhellingverticaal} it is clear that the gradients in the areas in WENSS LSS  where WENSS is relatively bright or faint have very different distributions: whereas the WENSS faint areas show no clear gradients, the WENSS bright areas show a clear bimodal distribution of gradients at about +2 radians/deg and about $-$2 radians/deg over a wide range of Galactic longitudes. Furthermore this bimodal distribution is present for all the interval lengths we probed (from 0.7\degr\hspace{0.2mm} to 5.6\degr). 
This last fact leads us to conclude that polarization-angle structure on scales that are small compared to the interval length do not dominate our gradient-fitting analysis (these structures only increase the $\chi^2$ of the fit), and that we are indeed sampling an underlying large-scale distribution.



\subsection{Gradients along Galactic longitude}\label{gradlon}
\begin{figure}[t]
\resizebox{\hsize}{!}{\includegraphics{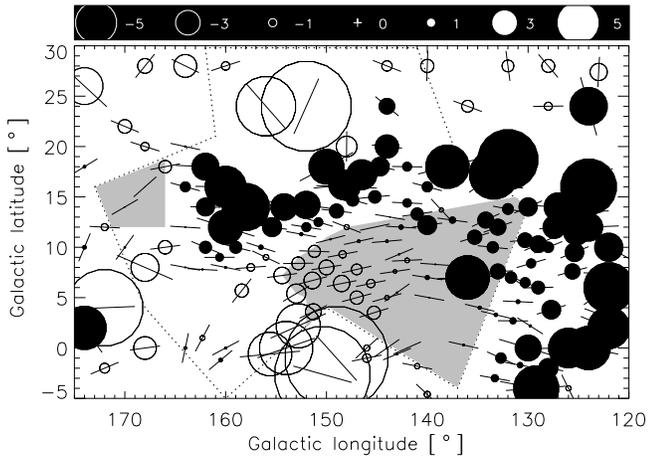}}
\caption{$RM$ and direction of \vec{B}$_{\perp}$ derived from BS polarization-angle measurements in the area covered by WENSS (enclosed by the dotted line) and its surroundings. The BS bright areas are shown as grey filled polygons. $RM$ are shown when the fit to the observed polarization angles has $\chi^2_{\mathrm{red}}<5$, and polarization angles should be available for at least 4 of the 5 wavelengths. The scale of the plotted symbols is linear and is illustrated in the top panel, units are radians/m$^2$. 
}
\label{rmbs}
\end{figure}

Gradients along Galactic longitude
${\partial\phi_{\mathrm{pol}}}/{\partial l}$ behave different from
gradients along Galactic latitude. In Fig. \ref{slopeshor} we show the
average gradient per strip along Galactic longitude. We only plotted
strips that contain at least 4 usable gradients. The weighted average
of the datapoints is indicated by a solid line. Since there is no
bimodal distribution of the gradients we did not have to separate the
fitted gradients as we did for gradients along Galactic latitude. From
this figure it is clear that the average gradient per strip is not
significantly different from 0 radians/deg. In the original WENSS
dataset we found the same result for gradients along Galactic
longitude.
The increased spread of the average gradient per strip and also the
increased errors per datapoint above a Galactic latitude of $\approx$
18\degr\hspace{0.2mm} coincides with a change in scale of the structure in
polarization angle, which we noted earlier in the WENSS LSS mosaic in
Fig. \ref{6p_phi_wenss_wensslss}, and/or a decrease in signal-to-noise ratio.

For the different interval lengths that we probed for strips along
Galactic longitude (1.4\degr, 2.1\degr, 2.8\degr, 4.2\degr\hspace{0.2mm} and
5.6\degr) the distribution of the gradients remains single-peaked, and
the average gradient along Galactic longitude is always close to 0 radians/deg,
but the average $\chi^2_\mathrm{red}$ per gradient increases as the
interval length increases from $<$ 5 to about 10 for the longest
intervals.

\section{Comparison to previous results}\label{comparison to previous results}

\begin{figure}[t]
\resizebox{\hsize}{!}{\includegraphics{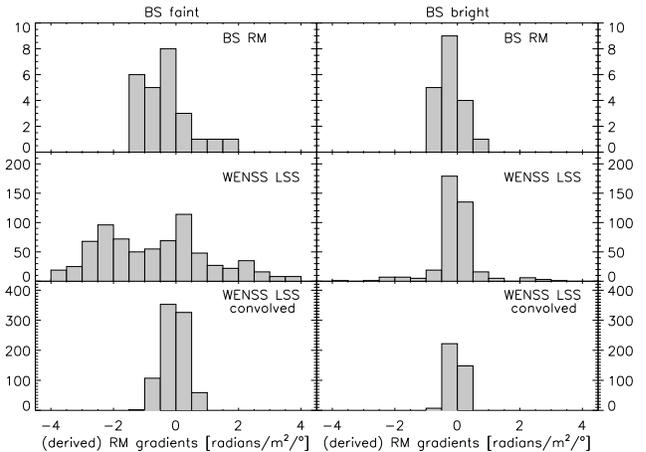}}
\caption{Histograms of the gradients fitted to the $RM$ we derived from the BS polarization-angle measurements (top row), and the gradients $(1/\lambda^2)\cdot\partial\phi_{\mathrm{pol}}/\partial b$ derived from WENSS LSS (bottom two rows). The $RM$ gradients we derive only use $RM$ with $\chi^2_{\mathrm{red}}<5$, and polarization angles should be available for at least 4 of the 5 wavelengths. The bottom row shows gradients in the WENSS LSS data after we convolved it with a 2.3\degr\hspace{0.2mm} beam. Gradients in the left column were fitted to the BS faint data, gradients in the right column were fitted to the BS bright data. The gradients fitted to the BS $RM$ were at most 5\degr\hspace{0.2mm} long, the gradients fitted to the WENSS LSS data were 4.2\degr\hspace{0.2mm} long. See the text for details.
}
\label{comp_gradients}
\end{figure}

We assume that the amount of Faraday rotation is given by $\phi-\phi_0 = RM\lambda^2 $. 
This means that if there are no gradients in the intrinsic polarization angle of the emitted radiation $\phi_0$,
the gradients in polarization angle $\partial\phi_{\mathrm{pol}}/\partial x$
(where `$x$' is either Galactic longitude or latitude) can be directly
translated into gradients in $RM$ with:

 \begin{equation}
   \frac{\partial RM}{\partial x}=\frac{1}{\lambda^2}\frac{\partial\phi_{\mathrm{pol}}}{\partial x} \label{translation}
 \label{rm_grad}
 \end{equation}

\noindent
$\phi_0$ varies very slowly with position in the region we are considering (see the directions of \vec{B}$_{\perp}$ plotted in Fig. \ref{rmbs}), which means that the contribution of $\partial \phi_0/\partial x$ to $\partial RM/\partial x$ is small. When we include the measured variations in $\phi_0$ in our analysis we find identical results.

To test the reliability of our `translation' of polarization-angle gradients as $RM$ gradients we calculated gradients in $RM$ based on the BS dataset. We rederived the $RM$ from the BS polarization-angle measurements to get also a $\chi^2_\mathrm{red}$ measure of the quality of the fitted $RM$, using the 1/signal-to-noise estimate of the error in the measured polarization angle described in Brouw \& Spoelstra (\cite{brouwspoelstra76}). In calculating $RM$ gradients we only considered $RM$ with $\chi^2_{\mathrm{red}}<5$, based on polarization angles for at least 4 of the 5 observing frequencies. In Fig. \ref{rmbs} we show these $RM$. 
 The $RM$ gradients we calculated from the BS data use pairs of $RM$ for which the separation in Galactic longitude was smaller than 0.2 times the separation in Galactic latitude to constrain the orientation of the gradients fitted to the $RM$. 
We only fitted gradients to lines-of-sight that were less than 5\degr\hspace{0.2mm} apart, which is slightly larger than the average distance between the datapoints shown in Fig. \ref{rmbs}. 



Spoelstra (\cite{spoelstra84}) points out that for a given sampling of $\lambda^2$ space the smallest $\lambda^2$ distance between two consecutive datapoints will set a maximum to both the $RM$ that can be determined and to the $B_{\|}$ that is derived from this $RM$. Spoelstra's argument is however based on only 2 datapoints, and the other available datapoints will make this criterion less strict. 
The $RM$ that were calculated by Haverkorn et al. (\cite{haverkorn03a}) are comparable in size to the $RM$ found by BS in the same region, but are derived from 5 MHz wide bands around about 350 MHz. An additional 180\degr\hspace{0.2mm} difference in polarization angles between two such consecutive bands would require a $RM$ that is about 150 radians/m$^2$ larger than they derive. In the direction we are considering such large $RM$ are only observed in pulsars that lie in or beyond the Perseus arm, and since depolarization effects will limit the line-of-sight, we do not believe that such large $RM$ should be observable in these data. Therefore we think that also the BS $RM$ are correct and do not suffer from `aliasing' due to the poor $\lambda^2$ sampling.

In Fig. \ref{comp_gradients} we show the gradients we fitted to the BS $RM$ separately for the BS faint and bright areas. If we found a significant difference between the polarization-angle distribution in WENSS LSS and the original distribution in WENSS such an area is identified as BS bright. In Fig. \ref{rmbs} we indicated the BS bright areas by a grey shading. 
In the second row of Fig. \ref{comp_gradients} we show the distribution of the gradients we fitted to the polarization-angle data, scaled to gradients in $RM$ by the factor of $1/\lambda^2$ in Eqn. \ref{translation}. These gradients were fitted over an interval of 4.2\degr. 

To see if the distribution of the BS $RM$ gradients can be scaled up to the gradients we derived from WENSS LSS we used a Kolmogorov-Smirnov two sample test using Eqn. \ref{rm_grad} as a description for the mapping of the BS gradients onto the WENSS gradients, where we apply a scale factor to the BS $RM$. At the 5\% confidence level for rejection scale factors between 1.7 and about 3.1 are permitted, with an average value of about 2.4. The KS probability  has a maximum of 0.7 for a scale factor of 2.2. 

In the BS faint region the $RM$ gradients we derived are about a factor of 2.4 smaller than the $RM$ gradients implied by the WENSS data. To simulate the effects of the difference in size between the WENSS and BS beams we convolved the WENSS LSS map with the 2.3\degr\hspace{0.2mm} beam of the BS data at 408 MHz. The results for the convolved data are shown in the bottom row of Fig. \ref{comp_gradients}. Clearly the larger telescope beam of the BS data can account for the narrowing of the distribution and the apparent disappearance of the negative mode. 

\section{Physical picture}\label{secdiscussion}
In this section we will interpret our data in terms of a simple model
of the ISM. In Sect. \ref{model} and \ref{Bz} we will derive
information on the magnetic field component along the line-of-sight
and on the component perpendicular to the Galactic plane. In
Sect. \ref{fan} we will discuss what we infer from our data about the
so-called `fan' region.

In our model we simulate a single line-of-sight through a volume of ISM with synchrotron emission and/or Faraday rotation. This volume contains both a regular and a random magnetic field. The regular field has a constant strength and direction, and the random field has a constant strength, but it changes direction on a scale that is (much) smaller than the size of the simulated volume. This ensures that there are enough draws of the orientation of the random field component along the line-of-sight that its statistical properties can be calculated analytically. The amount of emission only depends on the length of the total magnetic field vector (the sum of the regular and random components) projected perpendicular to the line-of-sight, as indicated in Appendix A of Haverkorn et al. (\cite{haverkorn04}). Similarly, the total magnetic field projected along the line-of-sight sets the amount of Faraday rotation. However, only a fraction of the line-of-sight is filled with cells that contain Faraday rotating electrons; in each of these cells the electron density is constant. The simulated volume of ISM is illuminated from the back by polarized synchrotron radiation; this background is assumed to have no internal Faraday rotation. 

One important ingredient in this model is the dispersion measure $DM$, the line-of-sight integral of the electron density, $DM = \int\limits_{l.o.s.} n_e\hspace{0.2mm}\mbox{d}l$, [$DM$]=cm$^{-3}$pc, which we cannot determine from our observations. To estimate $DM$ we integrated the NE2001 electron density model by Cordes and Lazio (\cite{cordeslazio}) out to  500 pc (`local') and 2 kpc (about the distance of the Perseus arm for these lines-of-sight). Since pulsars in the second Galactic quadrant that lie beyond the Perseus arm have $RM$ that are typically 100 radians/m$^2$, at least an order of magnitude larger than $RM$ of pulsars that lie in front of the arm (see the Galactic distribution of pulsars in e.g. Weisberg et al. \cite{weisberg04}), we argue that the magnitude of the BS $RM$, typically less than 10 radians/m$^2$, puts the Faraday-rotating electrons that produce these $RM$ between us and the Perseus arm, which means that the 2 kpc we use is a reasonable upper limit to the length of the line-of-sight. Note that for the 2 kpc line-of-sight and a Galactic latitude of 18\degr\hspace{0.2mm} (the maximum latitude that we will probe - see Table \ref{estimated_rm_bs_av}), the line-of-sight reaches a maximum height above the Galactic plane of about 600 pc, which means that we do not reach out into the Galactic halo.

\subsection{Deriving the line-of-sight magnetic field component}\label{model}
In this section we will derive strengths for the magnetic field component along the line-of-sight for a number of Galactic latitudes. We will use the $RM$ determined from the BS data and the $DM$ we calculated from the NE2001 model for the 2 assumed lengths of the line-of-sight.

Following Sokoloff et al. (\cite{sok98}) we introduce the `intrinsic rotation measure' $\mathcal{R}$ defined as the line-of-sight integral of the product of the magnetic field and the electron density $\mathcal{R} = 0.81\int\limits_{l.o.s.} n_e\vec{B}\cdot\mbox{d}\vec{l}$, [$B$]=$\mu$G, [$n_e$]=cm$^{-3}$, [$|\vec{l}|$]=pc, and [$\mathcal{R}$]=radians/m$^2$. As is customary the length of a vector will be denoted by $B = |\vec{B}|$. Since it is an integral along the line-of-sight it will not depend on the `clumpiness' of the distribution of the Faraday rotating electrons. 

If the line-of-sight is long enough, or if the average is taken over a number of lines-of-sight that are short, the contribution of the random magnetic field to $\mathcal{R}$ will cancel out in our model, and $\mathcal{R} \propto DM\hspace{0.2mm}B_{\mathrm{reg},\|}$. This means that we can calculate $B_\mathrm{reg,\|}$ from $\mathcal{R}$ and $DM$ using

 \begin{equation}
   B_\mathrm{reg,\|} = \frac{\mathcal{R}}{0.81\hspace{0.2mm} DM} = \frac{\int\limits_{l.o.s.}^{observer} n_e\vec{B}\cdot\mbox{d}\vec{l}}{\int\limits_{l.o.s.} n_e \mbox{d}l} \equiv \langle B_{\mathrm{reg},\|}\rangle_{n_e}
 \label{Bparr}
 \end{equation}

\noindent
where the magnetic field strength is in $\mu$G. The ratio of integrals on the r.h.s. of Eqn. \ref{Bparr} defines the electron density weigthed average $B_{\mathrm{reg},\|}$, or $\langle B_{\mathrm{reg},\|}\rangle_{n_e}$. Since the $B_\mathrm{reg,\|}$ we calculate from Eqn. \ref{Bparr} is defined to be constant throughout our model, $\langle B_{\mathrm{reg},\|}\rangle_{n_e}$ will be equal to $B_\mathrm{reg,\|}$ in this case. 

The rotation measure $RM$ is defined as the derivitive of polarization angle with respect to wavelength squared and, contrary to $\mathcal{R}$, it does depend on the exact distribution of emitting and Faraday rotating regions along the line-of-sight. This has been shown for example in Sokoloff et al. (\cite{sok98}). Therefore the ratio of $RM$ to $\mathcal{R}$ will be different for different configurations of emitting and Faraday rotating regions.

We use BS polarization-angle measurements with Galactic longitudes between about 150\degr\hspace{0.2mm} and 160\degr\hspace{0.2mm} and Galactic latitudes between 5\degr\hspace{0.2mm} and 20\degr, excluding the Auriga region between 156\degr\hspace{0.2mm} $\lesssim$ $l$ $\lesssim$ 165\degr\hspace{0.2mm} and 12\degr\hspace{0.2mm} $\lesssim$ $b$ $\lesssim$ 18\degr\hspace{0.2mm} where the $RM$ might be affected by local structure. Most of the BS data we consider thus lie in an area of WENSS LSS where LSS does not dominate. We estimated $RM$ and the intrinsic polarization angle of the emitted radiation $\phi_0$ from neighbouring (measured) lines-of-sight along 6 Galactic latitudes at a Galactic longitude of 152\degr, as shown in the first 3 columns of Table \ref{estimated_rm_bs_av}, and the errors given in the headings of these columns reflect our estimate of the spread around these averages. $\phi_0$ can be derived by extrapolating the BS polarization angles to $\lambda$=0 meters. Note that in the 3rd column of Table \ref{estimated_rm_bs_av} we indicate the direction of the magnetic field perpendicular to the line-of-sight, $\hat{\bf{B}}_{\perp}$, which is perpendicular to the direction indicated by $\phi_0$. At other Galactic longitudes the number of reliable $RM$ is either much lower, or LSS is missing from the interferometer observations. To correct the $RM$ in Table \ref{estimated_rm_bs_av} for beam effects these values had to be multiplied by a factor of 2.4 $\pm$ 0.7 as discussed in Sect. \ref{comparison to previous results}. 

In columns 4 and 6 of Table \ref{estimated_rm_bs_av} we show the values for $B_{\mathrm{reg},\|}$ we calculated from Eqn. \ref{Bparr} for the two assumed lengths of the line-of-sight. The error in $B_{\mathrm{reg},\|}$ is based on the range in Galactic latitude of the position of the line-of-sight (which gives a range in allowed $DM$), the error in the estimated $RM$, and the uncertainty in the scale factor between the WENSS and BS $RM$ gradients.
We compared the $B_{\mathrm{reg},\|}$ found using Eqn. \ref{Bparr} to the $B_{\mathrm{reg},\|}$ from the model by Haverkorn et al. (\cite{haverkorn04}), in the remainder referred to as `H04'. In their model Faraday rotation occurs in the synchrotron-emitting thin disc in cells filled with thermal electrons. Cell sizes range between 1 and 60 pc. These cells have a constant electron density of 0.08 cm$^{-3}$, and the filling factor assumed in this model is 0.2 (values based on Reynolds \cite{reynolds91}). The height of the synchrotron emitting thin disc is 180 pc (Beuermann et al. \cite{beuermann85}). The H04 model contains both a regular magnetic field, with constant strength and direction, and a random field, with constant strength, but a direction that is randomly drawn in each cell. The thin disc is illuminated from the back by a synchrotron-emitting halo. Emission and Faraday rotation in this halo are assumed to occur on such large angular scales that the emission from the halo has a uniform total and polarized intensity and direction of the plane of polarization.

The H04 model simulates a large number of lines-of-sight that each have their own distribution of Faraday rotating cells. By specifying the strength and the direction of \vec{B_{\mathrm{reg},\|}} and \vec{B_{\mathrm{reg},\perp}}, the strength of \vec{B_\mathrm{ran}}, and the total and polarized intensities of the background illuminating the thin disc, the H04 model predicts by numerical radiative transfer the observables Stokes I, $RM$, Stokes Q and U and their standard deviations (the spread of the observables being created by the different configurations of Faraday rotating cells along the line-of-sight). By comparing these output parameters to their observations, H04 constrain the range of allowed input parameters. 

One major advantage of the H04 model over our own analytical model is that the numerical treatment in the H04 model can predict $RM$. $RM$ is difficult to calculate analytically, since every configuration of Faraday rotating cells will have a different amount of depolarization, which means that the behaviour of the polarization angles with wavelength squared will be different. Therefore our model depends on a relation between $RM$ and $\mathcal{R}$ as input, which we inferred by comparing our results to the ones from the H04 model.

H04 find a $B_{\mathrm{reg},\|}$ of -0.42 $\mu$G in the direction of their Auriga field. When we use Eqn. \ref{Bparr} we find a value of -0.40 $\mu$G if we use the same input $DM$ as in the H04 model, and by setting $\mathcal{R}=RM$. From this we conclude that setting $\mathcal{R}=RM$ works well in this case.
The $DM$ used in the H04 model in this direction is 10 cm$^{-3}$pc, lower than the 31 cm$^{-3}$pc we determine from the NE2001 model by integrating it out to the same distance. 
In part this difference in $DM$ is due the filling factor that is used in the H04 model. For the clumpy disc they model, where all cells have the same electron density, a filling factor of 0.4 instead of the 0.2 used in the H04 model would agree better with the Reynolds data. This doubles the $DM$ in the H04 model, bringing it closer to the value predicted from the NE2001 model.

\begin{figure}[t]
\resizebox{8cm}{!}{\includegraphics{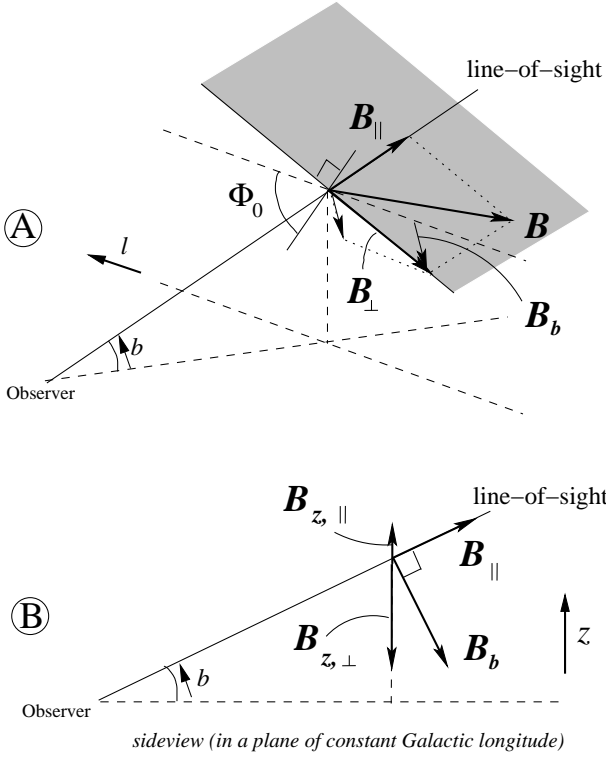}}
\caption{Scheme relating the different magnetic field components that are discussed in the text. Panel B shows panel A from the side in a plane of constant Galactic longitude. The magnetic field components along and perpendicular to the line-of-sight are \vec{B_\|}, resp. \vec{B_\perp}. The plane spanned by \vec{B_\|} and \vec{B_\perp} is indicated in grey. \vec{B_\perp} is directed perpendicular to $\phi_0$, both being defined in the plane perpendicular to the line-of-sight. \vec{B_b} is the component of \vec{B_\perp} along Galactic latitude. We indicated the directions of increasing Galactic longitude $l$ and latitude $b$ and height above the Galactic plane $z$, and defined $B_b$ and $B_z$ to increase when moving away from the Galactic plane.
}
\label{schetsrecon}
\end{figure}
 
\subsection{Reconstructing the 3D magnetic field vector. The magnetic field strength perpendicular to the Galactic plane}\label{Bz}
In this section we will derive information on the magnetic field component perpendicular to the Galactic plane \vec{B_\mathrm{z}}. Since our derivation will only apply to the regular magnetic field we will drop the `reg' subscript when we indicate field components. The relations between the different magnetic field components are illustrated in the top panel of Fig. \ref{schetsrecon}. First we will derive the strength of the magnetic field component perpendicular to the line-of-sight \vec{B_{\perp}} and its orientation. \vec{B_{\perp}} can be decomposed into a component along Galactic longitude and a component along Galactic latitude \vec{B_\mathrm{b}}. The component of \vec{B_{\perp}} along Galactic longitude does not contribute to \vec{B_\mathrm{z}}.
\vec{B_\mathrm{z}} can then be calculated from the projected \vec{B_\mathrm{b}} and \vec{B_\|}: $B_z=B_{z,{\perp}}+B_{z,{\|}}=B_b\cos(b)-B_{\|}\sin(|b|)$, as illustrated in the bottom panel in Fig. \ref{schetsrecon}. The sign of $B_z$ is such that a positive $B_z$ points away from the Galactic plane both for positive and negative Galactic latitudes. The sign of $B_b$ should be chosen accordingly. 

The strength of \vec{B_{\perp}} cannot be determined from our data. However, if we interpret the different $B_\|$ that we determine in Table \ref{estimated_rm_bs_av} as the result of different orientations of \vec{B} with respect to the line-of-sight (assuming that \vec{B} has a constant length for these lines-of-sight), $B_{\perp}$ can be calculated as a projection of \vec{B}. To do this $B$ has to be known. In the vicinity of the sun pulsar $RM$ give a $B$ that is typically about 2 $\mu$G, and equipartition arguments indicate for the total magnetic field (regular+random) a strength of about 6.5 $\mu$G (Beck et al. \cite{beck96}). In H04 both $B_{\|}$ and $B_{\perp}$ are derived, which give a $B \approx 3.3$ $\mu$G. If we assume that $B$ in the direction of H04's Auriga field is the same as in our data, we only have to rescale their $B$ to the lengths of the lines-of-sight we use. The scaling of $B_\|$ depends only on $DM$, and we can use our analytical model to derive the scaling properties of $B_\perp$ from Stokes I and $\sigma_\mathrm{I}$, in a similar way as was used in H04. In this way we find $B$ = 3.2 $\pm$ 0.5 and 2.0 $\pm$ 0.4 $\mu$G for our 500 pc and 2 kpc lines-of-sight.

\begin{table*}[t]
\centering
\caption[]{Estimated average $RM$ and direction of \vec{B}$_{\perp}$ (`{$\hat{\bf{B}}_{\perp}$}') from the BS data at a longitude of about 152\degr\hspace{0.2mm} for different Galactic latitudes. Estimates of the uncertainties in the parameters are given in the fourth row and in the columns for $B_{\mathrm{reg},\|}$, and are based on the range in Galactic latitude of the position of the line-of-sight, the spread of both $RM$ and direction of $\hat{\bf{B}}_{\perp}$, the uncertainty in the scale factor between the WENSS and BS $RM$ gradients, and the modeling uncertainties in the H04 model. Using Eqn. \ref{Bparr} the measured $RM$ (after being corrected for beam effects as described in Sect. \ref{comparison to previous results}) together with $DM$ calculated from the NE2001 model out to 500 pc and 2 kpc, can be used to calculate the magnetic field component along the line-of-sight, $B_{\mathrm{reg},\|}$. $B_z$ can then be calculated by projecting \vec{B_{\|}} and the component of \vec{B_\perp} along Galactic latitude, as is illustrated in the bottom panel of Fig. \ref{schetsrecon}.
}
\begin{tabular}{rrr|rr|rr} \hline\hline
\noalign{\smallskip}
 & & & \multicolumn{2}{c|}{500 pc line-of-sight} & \multicolumn{2}{c}{2 kpc line-of-sight} \\
 $b$\hspace{1.4mm} & $\langle RM_{BS}\rangle$\hspace{-2.8mm} & $\hat{\bf{B}}_{\perp}$\hspace{-0.2mm} & 
 $B_{\mathrm{reg},\|}$\hspace{4.3mm} & $ B_z$\hspace{1.7mm} & 
 $B_{\mathrm{reg},\|}$\hspace{4.3mm} & $ B_z$\hspace{1.7mm} \\
 (\degr)\hspace{0.4mm} & (radians/m$^2$)\hspace{-5mm} & (\degr) & 
 ($\mu$G)\hspace{4.3mm} & ($\mu$G) & 
 ($\mu$G)\hspace{4.3mm} & ($\mu$G) \\
 0.5 & 0.3 & 6\degr & & 0.32 & & 0.20 \\
\noalign{\smallskip}
\hline
\noalign{\smallskip}
 \hspace{0.3mm}7$^{\star}$ & 2.0    &  10   &    0.24 $\pm$ 0.08 &    0.51 &    0.07 $\pm$ 0.02 &    0.33 \\
 9\hspace{1.3mm}           & 1.5    &   0   &    0.18 $\pm$ 0.06 & $-$0.03 &    0.06 $\pm$ 0.02 & $-$0.01 \\
 10.5$^{\dagger}$          & 0      &   -   &       0 $\pm$ 0.04 &       - &       0 $\pm$ 0.01 &       - \\
 12\hspace{1.3mm}          & $-$1.2 & $-$12 & $-$0.14 $\pm$ 0.05 & $-$0.61 & $-$0.05 $\pm$ 0.02 & $-$0.39 \\
 14\hspace{1.3mm}          & $-$3.4 & $-$14 & $-$0.41 $\pm$ 0.12 & $-$0.64 & $-$0.14 $\pm$ 0.04 & $-$0.43 \\
 18$^{\dagger}$            & 0      &   -   &       0 $\pm$ 0.03 &       - &       0 $\pm$ 0.01 &       - \\
\noalign{\smallskip}
\hline\hline
\noalign{\smallskip}
\end{tabular}
\begin{list}{}{}
\item[$^{\star}$] the $b=7$\degr\hspace{0.2mm} datapoint lies in a part where the BS data is `intermediate' in brightness, whereas all other datapoints are in BS faint areas. The $B$ could therefore be different compared to other datapoints.
\item[$\dagger$] at these latitudes $\phi_0$ could not be reliably estimated, which means that $B_z$ could not be determined.
\end{list}
\label{estimated_rm_bs_av}
\end{table*}

$B_b$ can be derived from  $B_b=B_{\perp}\sin(2(\phi_0+\pi/2))$. Note that since $\phi_0$ follows from 1/2 arctan(U/Q), where U and Q refer to Stokes U and Q resp., it has a period of 180\degr, which is why there is a factor of 2 in this equation. The plane of polarization (and therefore also the direction of \vec{B_\perp}) is perpendicular to $\phi_0$, which accounts for the $\pi/2$ in the expression for $B_b$. Once the orientation of the plane of polarization is known, \vec{B_{\perp}} can still point in 2 directions. The global geometry of the magnetic field can be used to determine in which of the two possible directions \vec{B_{\perp}} is pointing: Brown et al. (\cite{brown03}) and Johnston-Hollitt et al. (\cite{hollitt04}) derive from $RM$ of extragalactic radiosources that the field points towards smaller Galactic longitudes. We will assume that the features we observe in WENSS LSS follow the same global field direction.

The errors we quote for $B_\|$ and $B_z$ are based on the range in Galactic latitude for the position of the line-of-sight, the spread in $RM$ and the orientation of $\hat{\bf{B}}_{\perp}$ and (in the case of $B_z$) also uncertainties in the parameters modeled by H04.
Even though our estimates of $B_z$ are 2$\sigma$ values at best, $B_z$ is between 1/2 to 4 times $B_\|$, for both lengths of the line-of-sight that we use. Since in all these cases the $B_\|$ and $B_b$ components are small compared to the total $B$, most of \vec{B} should be oriented along Galactic longitude, and therefore the magnetic field must be azimuthal, and directed towards smaller Galactic longitudes. Using pulsar observations Han and Qiao (\cite{hanqiao}) estimate that $B_z \approx 0.2-0.3$ $\mu$G, similar to our results.

To reach these conclusions we made a number of assumptions that, though reasonable, are hard to prove using our current data. 
We did not take into account the intrinsic difficulty of constraining the NE2001 model in the second Galactic quadrant which is the result of the small number of known pulsars in this direction. 
Since this only influences $DM$ and therefore $B_\|$, but does not influence $B_\perp$, we think 
it unlikely that the error in $DM$ is so large that it influences our conclusion of an azimuthal field. With better datasets, of the type described in Haverkorn et al. (\cite{haverkorn03a} and \cite{haverkorn03b}), some of these assumptions could be tested for their validity, and the errors in $B_\|$ and $B_z$ could also be reduced. We have obtained several of such datasets recently. 

To investigate if the results we present here can be considered `typical' for this quadrant we can look at the BS $RM$ that lie outside the area covered by WENSS, since we showed in Sect. \ref{comparison to previous results} that the BS $RM$ can be considered as scaled-down versions of WENSS $RM$ (at least in the BS faint part of WENSS LSS). Outside WENSS the BS sampling however becomes much poorer, and also the $RM$ seem to become much less correlated (which is probably in part due to the poorer sampling). Therefore it is at the moment not possible to investigate whether the results we present here apply only to a local feature or are more general. 




\subsection{Implications for the fan region}\label{fan}
The region between 136\degr\hspace{0.2mm} $\lesssim$ $l$ $\lesssim$ 152\degr\hspace{0.2mm} and $-$3\degr\hspace{0.2mm} $\lesssim$ $b$ $\lesssim$ 13\degr\hspace{0.2mm} is part of the so-called `fan' region that we already mentioned in Sect. \ref{secresults}. In the original WENSS dataset (without the single-dish information) the polarization-angle gradients in this area are bimodally distributed like in the rest of WENSS, as is also clear from the bottom panel in Fig. \ref{hist_slopes_run44_run45_run92}, where there is a lack of gradients around 0 radians/m$^2$/deg even though this figure includes polarization-angle gradients in the fan region.
One difference between the fan and the region outside the fan is that the ratio of positive to negative polarization-angle gradients in WENSS inside the fan is about 1:1, whereas this ratio is more 1:3 for WENSS as a whole.


We suggest that the fan is a region of enhanced emission that lies in front of the emission we see in WENSS. Due to the short lines-of-sight to the fan in this model, Faraday rotation effects, in particular variations in the amount of Faraday rotation across the fan surface, will be small. The polarized signal emitted by the fan will then be modulated by the foreground Faraday screen on angular scales that are too large to be detected by an interferometer due to its missing short spacings. The longer lines-of-sight to the background emission would make Faraday rotation effects strong enough that the background \emph{can} be detected by the interferometer. We therefore identify structure we see in WENSS as resulting from the Faraday modulation of this background emission. If the fan would lie at the same distance as the WENSS emission it should also be visible in the interferometer data. In the large single-dish beam the background emission, which is highly structured in polarization angle due to Faraday rotation, will get beam depolarized. If the fan would lie at the same distance as the background it too would get beam depolarized in the single-dish data, contrary to what we observe.

In this model the separation along the line-of-sight between the fan and the background emission is determined by the amount of small-scale structure in polarization angle. If the amount of structure on small angular scales is smaller than what we derive from WENSS (e.g. due to missing short baselines) the separation in distance becomes less. However the similarity of BS $RM$ inside and outside the fan, supplemented by the presence of a bimodal WENSS distributed in both these regions, suggests that we are not grossly overestimating the variation of polarization angles on small scales.


Spoelstra (\cite{spoelstra84}) concludes from a comparison between the polarization angles of starlight emitted by stars at known distances and $\phi_0$ that the Faraday rotating medium responsible for the $RM$ observed towards the fan should lie within a few hundred parsecs. 
$RM$ derived from the diffuse ISM in the direction of the fan are small compared to $RM$ derived for pulsars that, in this part of the Galaxy, lie beyond the Perseus arm, which would also indicate that most of the Faraday rotating medium seen towards the fan lies in front of the Perseus arm.
Wolleben (\cite{wolleben05}) suggests that the source of the emission seen towards the fan region forms a single physical structure that extends both above and below the Galactic plane. He tentatively identifies the low polarized intensities near the Galactic plane that appear to separate the parts of the fan region above and below the Galactic plane as depolarization occuring in HII complexes. This puts the source of the emission observed towards the fan beyond these complexes. One of the furthest of these complexes (with a moderate correlation with polarized intensity) is the IC1795/1805 complex around ($l,b$)=(135\degr,1\degr) at 2.1 kpc. The observations by Wolleben et al. were carried out at 1.4 GHz, whereas Spoelstra uses also data at much lower frequencies. Since depolarization (both line-of-sight and beam) will be more important at lower frequencies, the bulk of the observed signal will be coming from closer to the observer than at higher frequencies, which makes comparing data at such different frequencies more difficult.
 
The fan region can clearly be seen in the $RM$ in Fig. \ref{rmbs}. One striking feature appears to be a (smooth) change in $RM$ from positive at around ($l,b$)=(150\degr,6\degr), to about 0 at $l$=142\degr\hspace{0.2mm} to negative at around ($l,b$)=(130\degr,11\degr). Furthermore the negative and positive $RM$ are of about the same magnitude. A physical model of this structure would require knowledge of the strength of the magnetic field component in the plane of the sky. The observations and modelling by Haverkorn et al. (\cite{haverkorn03b}) imply a strong ($\approx3.2$ $\mu$G) field in the plane of the sky, but their modelling includes also the `ring' feature at ($l,b$)=(137\degr,7\degr). There are a number of indications that this ring structure could be a different type of object than its surroundings : the single large $RM$ in Fig. \ref{rmbs} lies at the position of the ring, whereas the other $RM$ in its vicinity are smaller, and furthermore the single-dish data show a local depression at the position of the ring in the generally high polarized intensities seen towards the fan region. 

\section{Summary \& Conclusions}
We have studied the polarization properties of the diffuse Galactic radio background at 325 MHz in the region 130\degr\hspace{0.2mm} $\lesssim$ $l$ $\lesssim$ 173\degr, $-$5\degr\hspace{0.2mm} $\lesssim$ $b$ $\lesssim$ 31\degr\hspace{0.2mm} using the WENSS dataset. We determined gradients in polarization angle (which can be translated into gradients in $RM$), taking into account the periodic nature of the polarization-angle data. The WENSS data show gradients that are coherent over large areas. We find a bimodal distribution of gradients along Galactic latitude, with peaks at +2.1 radians/deg and $-$2.1 radians/deg, with about three times as many gradients in the negative mode as there are in the positive mode, and with hardly any gradients in between the two modes. Along Galactic longitude the gradients are in general nearly zero.

We investigated the importance of LSS missing in our data from the higher-frequency single-dish data obtained by Brouw\&Spoelstra (\cite{brouwspoelstra76}). We first interpolated between the lines-of-sight of the single-dish data, followed by an extrapolation to 325 MHz, the observing frequency of WENSS. We ran several consistency tests between the different frequencies present in the single-dish observations which show that the data allow this handling in a robust way, independent of the exact parameters we use. Our reconstruction of the LSS missing from WENSS is only statistical in nature, we do not claim to reconstruct the correct observables for individual pixels. We refer to the combined WENSS/BS datasets as `WENSS LSS'. We estimate that the combined dataset is still insensitive to structure on spatial scales between $\approx$1.5\degr\hspace{0.2mm} and $\approx$5\degr. 

After adding our estimate of missing LSS to the WENSS data, many polarization-angle gradients are `converted' from the steep bimodal WENSS distribution into a peak around zero radians/deg. The negative mode of WENSS gradients remains clearly visible. Since we constructed WENSS LSS as the (polarization) vector sum of WENSS and BS, the original WENSS or BS data will dominate if its polarization vectors are much longer than the vectors in the other dataset. Therefore in regions where the BS signal is strong the slowly-varying (nearly zero radians/deg) BS gradients will dominate WENSS LSS.

By using a Kolmogorov-Smirnov (KS) test to compare the polarization-angle gradients from the BS faint part of WENSS LSS to scaled-up gradients from the BS $RM$. We found that these distributions are similar, and we set limits to the scale factor using this KS test. We illustrated the importance of the difference in size between the WENSS and BS beams by convolving the WENSS LSS data with the BS 408 MHz single-dish beam, which made the clear bimodal distribution of polarization-angle gradients in WENSS LSS look more like the distribution of BS $RM$ gradients. 

 We proceeded by deriving the strength of the magnetic field component along the line-of-sight using these scaled-up $RM$ together with dispersion measures from the NE2001 thermal electron density model by Cordes \& Lazio (\cite{cordeslazio}), see Table \ref{estimated_rm_bs_av} where we show the values derived for lines-of-sight of lengths 0.5 and 2 kpc. By comparing the magnitude of the scaled-up $RM$ to $RM$ measured for pulsars in front of and beyond the Perseus arm we conclude that the Faraday rotation effects we are observing in the BS faint part of our data should be occuring between us and the Perseus arm.

We used the estimate of the strength of the total (regular) magnetic field by Haverkorn et al. (\cite{haverkorn04}) to derive the strength of the magnetic field in the plane of the sky \vec{B_\perp}, and even though our results are about 1$-$3$\sigma$ at best, we showed that the component of \vec{B_\perp} along Galactic longitude is much stronger than both the component of \vec{B_\perp} along Galactic latitude and \vec{B_\|}. We were also able to determine the strength of the magnetic field component perpendicular to the Galactic plane, $B_z$, shown in Table \ref{estimated_rm_bs_av}. 


Part of the area covered by WENSS LSS includes the `fan' region, an area bright in polarized intensity, but at the same time extended on large angular scales, which meant that it could not be detected by the WSRT. We note that the large-scale emission that is detected by the single-dish does not vary on the same angular scales as the WENSS data, and we suggest that the large-scale emission seen towards the fan lies in front of the emission we see in WENSS. The magnitude of the $RM$ observed towards the fan are small, indicating that they are caused by Faraday rotation occurring between us and the Perseus arm.

\begin{acknowledgements}
We would like to thank Titus Spoelstra and Wolfgang Reich for
supplying us with the Brouw \& Spoelstra '76 dataset. We would also
like to thank Maik Wolleben for useful discussion on the nature of the
fan region. Finally we would like to thank the anonymous referee for
constructive remarks that helped to improve the manuscript. The
Westerbork Synthesis Radio Telescope is operated by the Netherlands
Foundation for Research in Astronomy (NFRA) with financial support
from the Netherlands Organization for Scientific Research (NWO). MH
acknowledges support from the National Radio Astronomy Observatory,
which is operated by Associated Universities, Inc., under cooperative
agreement with the National Science Foundation.
\end{acknowledgements}

\appendix
\section{A new method for fitting linear gradients to periodic data}\label{appendixA}
The main problem when fitting functions to periodic data like polarization angles is the correct handling of the 180\degr\hspace{0.2mm} ambiguity. Finding the best-fitting slope and offset requires at the same time finding the number of times 180\degr\hspace{0.2mm} has to be added to each datapoint to minimize the distance between the line fitted to the data and the datapoint. This means that for $n$ datapoints $n+2$ degrees of freedom have to be solved, 1 degree of freedom per datapoint to determine the number of 180\degr\hspace{0.2mm} `wraps' ($N_i$; i=1,...,n), and 2 degrees to determine slope and offset. The only solution to this problem is that at least 2 of the $N_i$ are known a priori. The best-fitting slope and offset will however only depend on the (small number of) datapoints for which the $N_i$ are known, which can be shown by working out the equations that minimize $\chi^2$.




An alternative approach which takes the behaviour of all datapoints into account when determining slope and offset can be formulated as follows.
When making a linear fit one can pick a combination of slope and offset and then bring each datapoint to within 90\degr\hspace{0.2mm} of the imposed linear fit, which will minimize the $\chi^2$ for that fit. By comparing the $\chi^2$ for all possible combinations of slope and offset one then finds the best fit to the data. 

Instead of considering all possible slopes and offsets generally only combinations are considered that lie on a finite grid. For a periodic function only the central part of the complete (slope,offset) space has to be investigated. For polarization-angle gradients this means that only slopes between $-$90\degr/pixel and +90\degr/pixel and offsets between $-$90\degr\hspace{0.2mm} and +90\degr\hspace{0.2mm} have to be investigated since other slopes and offsets are indistinguishable from slopes and offsets in this central space.

\begin{figure}[t]
\resizebox{\hsize}{!}{\includegraphics{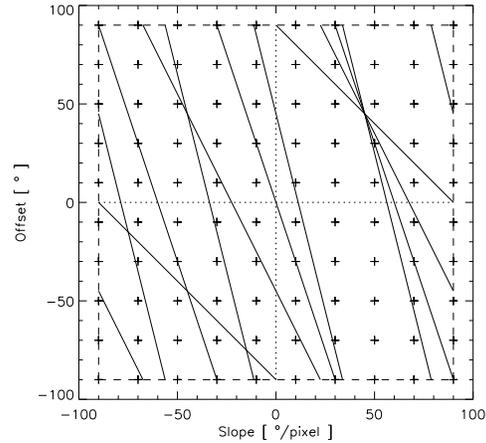}}
\caption{Illustration of which domains and configurations are allowed by a simple
example distribution of 4 pixels with coordinates (pixel,polarization angle)= (1,0\degr), (2,45\degr), (3,$-$90\degr) and (4,$-$45\degr). All domain boundaries that belong to a single datapoint run parallel to each other, and the boundaries are steeper if the datapoint is further from the origin of the fitted line, which is at pixel coordinate 0 in this example.
As a comparison we also show a grid that probes (slope,offset)
combinations (indicated as crosses) with grid spacing
20\degr/pixel$\times$20\degr.  }
\label{domains}
\end{figure}

In this appendix we discuss an alternative method which is not based on a grid of slopes and offsets. Instead it maps areas in (slope,offset) space (`domains') which share the same configurations ($N_1$,..., $N_n$) of 180\degr\hspace{0.2mm} wraps. For a given (slope,offset) combination the lowest $\chi^2$ is found when all datapoints are within 90\degr\hspace{0.2mm} of the fitted line formed by that (slope,offset) combination. Each domain boundary is thus formed by (slope,offset) combinations for which one datapoint is at exactly 90\degr\hspace{0.2mm} from the fitted line. Crossing this boundary means going to (slope,offset) combinations for which the datapoint will be further than 90\degr\hspace{0.2mm} from the fitted line, and it can be brought to within 90\degr\hspace{0.2mm} by adding an extra $\pm$180\degr\hspace{0.2mm} to the datapoint, which changes the configuration. Each datapoint thus defines a set of domain boundaries, and the ensemble of the domain boundaries of the individual datapoints will define the domains.

For each of these domains the best-fitting combination of slope and offset is found by a $\chi^2$ fit. Since all combinations of slopes and offsets in the domain share the same best fit, it becomes possible to cover the entire solution space by mapping the domains.

One possible cause of concern could be that the least-squares solution falls outside its domain. However the $\chi^2$ of this solution will always be higher than the minimum $\chi^2$ of the best fit in the domain the solution has crossed into: If the best-fitting (slope,offset) combination lies outside the domain it is supposed to be in, by definition of the domain borders 
at least one of the datapoints is further than 90\degr\hspace{0.2mm} from the line formed by that (slope,offset) combination. For that (slope,offset) combination it is however always possible to find a configuration where all datapoints are within 90\degr\hspace{0.2mm} of the line formed by that (slope,offset) combination, and this configuration will therefore give a better fit. 
This furthermore guarantees that the lowest $\chi^2$ in (slope,offset) space will always lie inside the domain it is supposed to be, and that by mapping the domains and using a $\chi^2$ fit in every domain we will find this best-fitting solution. 

The boundaries of the domains are set by the noise realization of the observations. Since the same distance between the datapoint and the linear fit is used to find the position of the domain borders and in the definition of $\chi^2$, the domain mapping approach will be no more sensitive to the influence of noise than a standard $\chi^2$ fit applied to a non-periodic dataset.

The configuration-mapping approach covers all solutions in (slope,offset) space and is therefore much more reliable than the grid-based approach which only probes the quality of certain (slope,offset) combinations. But furthermore it is very easy to figure out where the boundaries of the domains are, and therefore this method is also much faster. In one example of 4 datapoints at (pixel,polarization angle)= (1,0\degr), (2,45\degr), (3,$-$90\degr) and (4,$-$45\degr) the grid-based approach using a coarse grid of 20\degr/pixel$\times$20\degr\hspace{0.2mm} already requires 100 gridpoints, whereas there were only 21 domains in this configuration that needed to be investigated (see Fig. \ref{domains}).

Sarala and Jain (\cite{saralajain01}) discuss a different method to find the best-fitting gradient for a periodic dataset. Their analysis includes all wraps of the individual datapoints and using the appropriate statistics they derive a maximum-likelihood criterion. By bringing the datapoints to within 90\degr\hspace{0.2mm} of the fitted gradient our method can use the standard non-periodic statistics.

The method we describe here can also be used for fitting $RM$ to polarization-angle observations at different wavelengths, since this would mean solving $\phi=\mbox{a}\lambda^2+\mbox{b}$ for the parameters `a' and `b', which is identical to the $\phi=\mbox{a}x+\mbox{b}$ we solved for the spatial $RM$ gradients. One important difference however is that the regular spacing of the pixels we fitted $RM$ gradients to limits the range of slopes that give unique fits to the data: slopes steeper than 90\degr/pixel are indiscernible from slopes that are flatter. If the polarization angles are irregularly sampled in $\lambda^2$ space there are no such limits, which means that there are no boundaries that limit the range of $RM$ that have to be investigated. The regular spacing of datapoints also causes the degeneracy that steeper gradients can be fitted with the same quality by adding $\pi$ radians to the second datapoint, $2\pi$ to the third etc. For an irregular spacing this degeneracy is (at least partially) lifted.



\begin{thebibliography}{}

   \bibitem[1977]{baars77} Baars, J.W.M., Genzel, R., Pauliny-Toth, I.I.K., Witzel, A.  1977, A\&A, 61, 99

   \bibitem[1996]{beck96} Beck, R., Brandenburg, A, Moss, D., et al. 1996, ARA\&A, 34, 155

   \bibitem[2003]{beck03} Beck, R., Shukurov, A., Sokoloff, D., Wielebinski, R. 2003, A\&A, 411, 99

   \bibitem[1971]{berkhuijsen71} Berkhuijsen, E.M. 1971, A\&A, 14, 359

   \bibitem[1985]{beuermann85} Beuermann, K., Kanbach, G., and Berkhuijsen, E.M. 1985, A\&A, 153, 17

   \bibitem[1967]{binghamshakeshaft} Bingham, R.G., \& Shakeshaft, J.R. 1961, MNRAS, 136, 347

   \bibitem[1976]{brouwspoelstra76} Brouw, W.N., \& Spoelstra, T.A.Th. 1976, A\&A, 26, 129

   \bibitem[2003]{brown03} Brown, J.C., Taylor, A.R., Wielebinski, R., Mueller, P. 2003, ApJ, 592, L29

   \bibitem[2002]{cordeslazio} Cordes, J.M., Lazio, T.J.W. 2002, astro-ph 0207156, and \\ \hspace{3mm} http://astrosun2.astro.cornell.edu/~cordes/NE2001/~ 


   \bibitem[1996]{ham96} Hamaker, J.P., Bregman, J.D., Sault, R.J. 1996, A\&AS, 117, 137

   \bibitem[1994]{hanqiao} Han, J.L., \& Qiao, G.J. 1994, A\&A, 288, 759


   \bibitem[2004]{han04} Han, J.L., Ferriere, K., Manchester, R.N. 2004, ApJ, 610, 820

   \bibitem[2006]{han06} Han, J.L., Manchester, R.N., Lyne, A.G., et al. 2006, ApJ, 642, 868

   \bibitem[1981]{haslam81} Haslam, C.G.T., Klein, U., Salter, C.J., et al. 1981, A\&A, 100, 209

   \bibitem[2000]{haverkorn00} Haverkorn, M., Katgert, P., De Bruyn, A.G. 2000, A\&A, 356, L13 

   \bibitem[2003a]{haverkorn03a} Haverkorn, M., Katgert, P., De Bruyn, A.G.  2003a, A\&A, 403, 1031 

   \bibitem[2003b]{haverkorn03b} Haverkorn, M., Katgert, P., De Bruyn, A.G.  2003b, A\&A, 404, 233 

   \bibitem[2004]{haverkorn04} Haverkorn, M., Katgert, P., De Bruyn, A.G. 2004, A\&A, 427, 169 

   \bibitem[2006a]{haverkorn06igps} Haverkorn, M., Gaensler, B.M., McClure-Griffiths, N.M., et al. 2006a, ApJS, in press

   \bibitem[2006b]{haverkorn06} Haverkorn, M., Gaensler, B.M., Brown, J.C., et al. 2006b, ApJL, 637, L33

   \bibitem[2004]{hollitt04} Johnston-Hollitt, M., Hollitt, C.P., Ekers, R.D. 2004, Statistical Analysis of Extra-galactic Rotation Measures. In The Magnetized Interstellar Medium, Proceedings of the conference, ed. B. Uyaniker, W. Reich, \& R. Wielebinski, p. 13

   \bibitem[1987]{junkes87} Junkes, N., F\"{u}rst, E., and Reich, W. 1987, A\&AS, 69,451


   \bibitem[1989]{randkulkarni} Rand, R.J., \& Kulkarni, S.R. 1989, ApJ, 343, 760

   \bibitem[1997]{reich97} Reich, P., Reich, W., F\"urst, E. 1997, A\&AS, 126, 413

   \bibitem[1997]{rengelink97} Rengelink, R., Tang, Y., De Bruyn, A.G., et al. 1997, A\&AS, 124, 259

   \bibitem[1991]{reynolds91} Reynolds, R.J. 1991, ApJ, 372, L17

   \bibitem[2001]{saralajain01} Sarala, S., \& Jain, P. 2001, MNRAS, 328, 623

   \bibitem[1996]{sault96} Sault, R.J., Hamaker, J.P. Bregman, J.D. 1996, A\&AS, 117, 149

   \bibitem[1998]{sok98} Sokoloff, D.D., Bykov, A.A., Shukurov, A., et al. 1998, MNRAS, 299, 189 

   \bibitem[1984]{spoelstra84} Spoelstra, T.A.T. 1984, A\&A, 135, 238

   \bibitem[2002]{stanimirovic} Stanimirovic, S. 2002, Short-Spacings Correction from the Single-Dish Perspective. In Single-Dish Radio Astronomy: Techniques and Applications, ed. S. Stanimirovic, D. Altschuler, P. Goldsmith, et al., ASP Conference Proceedings, vol. 278, 375

   \bibitem[2003]{taylor03} Taylor, A.R., Gibson, S.J., Peracaula, M., et al. 2003, AJ, 125, 3145

   \bibitem[2003]{uyaniker03} Uyaniker, B., Landecker, T., Gray, A.D., Kothes, R. 2003, ApJ, 585, 785

   \bibitem[2004]{weisberg04} Weisberg, J. M., Cordes, J. M., Kuan, B., et al. 2004, ApJS, 150, 317

   \bibitem[1993]{wieringa93} Wieringa, M., De Bruyn, A.G., Jansen, D., Brouw, W.N., Katgert, P. 1993, A\&A, 268, 215
  
   \bibitem[2005]{wolleben05} Wolleben, M., Ph.D. thesis Rheinischen Friedrich-Wilhelms-Universi\\t\"at Bonn, 2005. Available online on ADS. 
                        
\end{thebibliography}
\end{document}